\documentclass[showpacs,nofootinbib,showkeys,eqsecnum,prd,aps,preprint]{revtex4-1}
\usepackage{amssymb}
\usepackage{amsmath}
\usepackage{amsfonts}
\usepackage{graphicx}
\usepackage{bm}
\usepackage[T1]{fontenc}
\usepackage{graphicx}

\begin{document}

\title{Vacuum instability in slowly varying electric fields}
\author{S.~P.~Gavrilov${}^{a,c}$}
\email{gavrilovsergeyp@yahoo.com, gavrilovsp@herzen.spb.ru}
\author{D.~M.~Gitman${}^{a,b,d}$}
\email{gitman@if.usp.br}
\date{\today }

\begin{abstract}
Nonperturbative methods have been well-developed for QED with the so-called $%
t$-electric potential steps. In this case a calculation technique is based
on the existence of specific exact solutions (\textrm{in} and \textrm{out}
solutions) of the Dirac equation. However, there are only few cases when
such solutions are known. Here, we demonstrate that for $t$-electric
potential steps slowly varying with time there exist physically reasonable
approximations that maintain the nonperturbative character of QED
calculations even in the absence of the exact solutions. Defining the slowly
varying regime in general terms, we can observe a universal character of
vacuum effects caused by \ a strong electric field.\ In the present article%
{\large , }we find\ universal approximate representations for the total
density of created pairs and{\large \ }vacuum mean values of the current
density and energy-momentum tensor that hold true for arbitrary $t$-electric
potential steps slowly varying with time. These representations do not
require knowledge of the corresponding solutions of the Dirac equation, they
have a form of simple functionals of a given slowly varying electric field.%
{\large \ }We establish relations of these representations with leading
terms of the derivative expansion approximation.{\large \ }These results
allow one to formulate some semiclassical approximations that are not
restricted by the smallness of differential mean numbers of created pairs.
\end{abstract}


\affiliation{$^{a}$Department of Physics, Tomsk State University, Lenin Prospekt 36, 634050, Tomsk, Russia\\
$^{b}$P. N. Lebedev Physical Institute, 53 Leninskiy prospect, 119991,
Moscow, Russia\\
${}^{c}$Department of General and Experimental Physics, Herzen State
Pedagogical University of Russia, Moyka embankment 48, 191186
St.~Petersburg, Russia\\
$^{d}$Institute of Physics, University of S\~{a}o Paulo, CP 66318, CEP
05315-970 S\~{a}o Paulo, SP, Brazil\\
}

\maketitle

\section{Introduction\label{S1}}

It is well known that in QED with strong electriclike external fields there
exists so-called vacuum instability due to real particle creation caused by
the external field. A number of publications, reviews, and books have been
devoted to the effect of particle creation itself and to developing
different nonperturbative calculation methods in theories with unstable
vacuum: analytical (semiclassical and based on exact solutions) and
numerical; see Refs. \cite{Nikis79,General,FGS} for a review.\textrm{\ }Most
semiclassical and numerical methods are applied to Schwinger's effective
action and related formulas \cite{Schwinger51} (see Ref.~\cite{Dunn04} for a
review), to calculate the probability for a vacuum to remain a vacuum.%
{\large \ }They are well grounded for not very strong electric fields, when
the probability for pair creation is exponentially small. There exists the
derivative expansion approximation method, which being applied to
Schwinger's effective action, allows one to treat effectively slowly varying
strong fields \cite{DunnH98,GusSh99}. However, it should be noted that the
probability for a vacuum to remain a vacuum contains only little information
about the time evolution of vacuum effects caused by strong electric field.
It can be seen that in some situations in astrophysics and condensed matter
the time evolution of vacuum effects caused by strong electric fields is of
significant interest;{\large \ }e.g., see Refs. \cite%
{General,GavGitY12,lewkowicz10,Van+etal10,grcr12-15}. In the case of strong
external fields{\large ,} nonperturbative methods have been well developed
for QED with two specific configurations of external backgrounds, namely for
the so-called $t$-electric potential steps \cite{FGS,Tcase,GavGT06} and $x$%
-electric potential steps \cite{Xcase}. In both cases the calculation
technique is based on the existence of specific exact solutions (\textrm{in}
and \textrm{out} solutions) of the Dirac equation. Under this condition, all
the probability amplitudes and mean values in the backgrounds under
consideration have some nonperturbative integral representations via these
\textrm{in} and \textrm{out} solutions. At present, only a few types of $t$-
and $x$-electric potential steps are known when such solutions are known; we
call these cases exactly solvable cases. In QED with $t$-electric potential
steps, exactly solvable cases that have real physical importance are the
Sauter-like electric field; the so-called $T$-constant electric field (a
uniform electric field which acts during a finite time interval $T$,
including the constant electric field when $T\rightarrow \infty )$; and
exponentially growing and decaying electric fields. Using the corresponding
exact solutions, different characteristics of quantum processes related to
the particle creation were calculated in detail; see \cite%
{GavGitY12,TcaseCal1,GavG96a,Nikis70a,DunnH98,TcaseCal2,GavG08,GG06-08,AdoGavGit14,AdoGavGit16}%
, respectively. And here we come to the question of whether there exist
physically reasonable approximations in QED with the above described strong
backgrounds that\textrm{\ }maintain nonperturbative calculations and allow
one to go beyond dealing with the existence of the exact solutions? In this
article, we demonstrate that such a possibility exist in the case of QED
with the $t$-electric potential steps slowly varying with time and similar
possibility in the case of QED with $x$-electric potential steps\ will be
presented in our next publication.

In Sec. \ref{S2}, we give a definition of slowly varying $t$-electric
potential steps and revised vacuum instability due to such backgrounds for
the existing exactly solvable cases.\textrm{\ }In Sec \ref{S3}, we stress
universal features of the vacuum instability in these examples. We derive
universal approximate representations for the total density of created pairs
and vacuum mean values of current density and energy-momentum tensor (EMT)
components that hold true for arbitrary $t$-electric potential step slowly
varying with time. These representations do not require knowledge of the
corresponding solutions of the Dirac equation, they have a form of simple
functionals of a given slowly varying electric field.{\large \ }We establish
relations of these representations with leading terms of derivative
expansion approximation. These results allow one to formulate some
semiclassical approximations that are not restricted by the smallness of
differential mean numbers of created pairs. In the Appendix \ref{ApA}, we
briefly describe a nonperturbative formulation of QED with $t$-electric
potential steps.

\section{Slowly varying $t$-electric potential steps, exactly solvable cases
\label{S2}}

We call $E(t)$ a slowly varying electric field on a time interval $\Delta t$
if the following condition holds true:%
\begin{equation}
\left\vert \frac{\overline{\dot{E}(t)}\Delta t}{\overline{E(t)}}\right\vert
\ll 1,\ \ \Delta t/\Delta t_{\mathrm{st}}^{\mathrm{m}}\gg 1,  \label{svf1}
\end{equation}%
where $\overline{E(t)}$ and $\overline{\dot{E}(t)}$ are mean values of $E(t)$
and $\dot{E}(t)$ on the time interval $\Delta t$, respectively, and $\Delta
t $ is significantly larger than the time scale $\Delta t_{\mathrm{st}}^{%
\mathrm{m}}$ which is
\begin{equation}
\;\Delta t_{\mathrm{st}}^{\mathrm{m}}=\Delta t_{\mathrm{st}}\max \left\{
1,m^{2}/e\overline{E(t)}\right\} ,\,\;\,\Delta t_{\mathrm{st}}=\left[ e%
\overline{E(t)}\right] ^{-1/2}.  \label{svf2}
\end{equation}

Note that the time scale $\Delta t_{\mathrm{st}}^{\mathrm{m}}$ appears in
Eq.~(\ref{svf1}) as the time scale when the perturbation theory with respect
to the electric field breaks down and the Schwinger (nonperturbative)
mechanism is primarily responsible for the pair creation. In what follows,
we show that that this condition{\Large \ }is sufficient. We are primarily
interested in strong electric fields, $m^{2}/e\overline{E(t)}\lesssim 1$. In
this case, the second inequality in Eq.~(\ref{svf1}) is simplified to the
form $\Delta t/\Delta t_{\mathrm{st}}\gg 1$, in which the mass $m$ is
absent. In such cases, the potential of the corresponding electric steps
hardly differs from the potential of a constant electric field,
\begin{equation}
U\left( t\right) =-eA_{x}\left( t\right) \approx U_{c}\left( t\right) =e%
\overline{E(t)}t+U_{0},  \label{4.39}
\end{equation}%
on the time interval $\Delta t$, where $U_{0}$ is a given constant. This
behavior is inherent for the fields of known exact solvable cases with
appropriate parameters, namely, the peak field, the $T$-constant electric
field, and the Sauter-like electric field.

The complete sets of solutions of the Dirac equation, given by Eq.~ (\ref%
{t4.10}), are determined by the functions $\ _{\zeta }\varphi _{n}\left(
t\right) $ and $\ ^{\zeta }\varphi _{n}\left( t\right) $ which play the role
of in and out solutions of positive ($\zeta =+$) and negative ($\zeta =+$)
energy as $t\rightarrow \pm \infty $, respectively (cf. the Appendix). We
assume that the electric field is directed along the axis $x$. We choose
that before time $t_{\mathrm{in}}$ and after time $t_{\mathrm{out}}$ these
solutions are states with a definite momentum $p=\left( p_{x},\mathbf{p}%
_{\bot }\right) $ (where the index $\perp $ stands for components that are
perpendicular to the electric field) and spin polarization $\sigma .$ Then
the complete set of quantum numbers is $n=(\mathbf{p},\sigma )$. {\large \ }%
The functions $\ _{\zeta }\varphi _{n}\left( t\right) $\ and $\ ^{\zeta
}\varphi _{n}\left( t\right) $ are known explicitly for the following
electric fields.

(i) The Sauter-like (or adiabatic or pulse) electric field and its vector
potential have the form%
\begin{equation}
E\left( t\right) =E_{0}\cosh ^{-2}\left( t/T_{\mathrm{S}}\right) \,,\ \
A_{x}\left( t\right) =-T_{\mathrm{S}}E_{0}\tanh \left( t/T_{\mathrm{S}%
}\right) \,.  \label{2.8}
\end{equation}%
where the parameter $T_{\mathrm{S}}>0$ sets the time scale. The functions $\
_{\zeta }\varphi _{n}\left( t\right) $ and $\ ^{\zeta }\varphi _{n}\left(
t\right) $ and the number of created pairs $N_{n}^{\mathrm{cr}}$ are given,
for example, in Ref.~\cite{GavG96a}. We have the case of a slowly varying
field if
\begin{equation}
\sqrt{eE_{0}}T_{\mathrm{S}}\gg \max \left( 1,m/\sqrt{eE_{0}}\right) \,.
\label{asy1}
\end{equation}%
In this case, the leading contribution to the total number of pairs\ created
from vacuum is formed in the range of $\left\vert p_{x}\right\vert <eE_{0}T_{%
\mathrm{S}}$ and small $\pi _{\perp }\ll eE_{0}T_{\mathrm{S}}$. In this
range the differential mean numbers of created pairs have approximately the
following form
\begin{equation}
N_{n}^{\mathrm{cr}}\approx N_{n}^{\mathrm{as}}=\exp \left\{ -\pi T_{\mathrm{S%
}}\left[ p_{0}\left( +\infty \right) +p_{0}\left( -\infty \right) -2eE_{0}T_{%
\mathrm{S}}\right] \right\} \,,  \label{asy2}
\end{equation}%
where $p_{0}\left( \pm \infty \right) $ are the energies given by Eq.~(\ref%
{t4.ba}) in the Appendix. This distribution has a maximum at $p_{x}=0$. \
This maximum coincides with the differential number of created pairs in a
constant electric field \cite{Nikis70a,Nikis79},
\begin{equation}
N_{n}^{\mathrm{cr}}\approx N_{n}^{0}=e^{-\pi \lambda _{0}},\ \ \lambda _{0}=%
\frac{\pi _{\bot }^{2}}{eE_{0}}\,,\ \ \pi _{\perp }=\sqrt{\mathbf{p}_{\perp
}^{2}+m^{2}}.  \label{asy4}
\end{equation}

(ii) The so-called $T$-constant electric field does not change within the
time interval $T$ and is zero outside of it,%
\begin{equation}
E\left( t\right) =\left\{
\begin{array}{l}
0\,,\ \ t\in \mathrm{I} \\
E_{0}\,,\ \ t\in \mathrm{II} \\
0\,,\ \ t\in \mathrm{III}%
\end{array}%
\right. \Longrightarrow A_{x}\left( t\right) =\left\{
\begin{array}{l}
-E_{0}t_{\mathrm{in}}\,,\ \ t\in \mathrm{I} \\
-E_{0}t\,,\ \ t\in \mathrm{II} \\
-E_{0}t_{\mathrm{out}}\,,\ \ t\in \mathrm{III}%
\end{array}%
\right. \,,  \label{t7}
\end{equation}%
where $\mathrm{I}$ denotes the in region $t\in \left( -\infty ,t_{\mathrm{in}%
}\right] $, $\mathrm{II}$ is the intermediate region where the electric
field is nonzero $t\in \left( t_{\mathrm{in}},t_{\mathrm{out}}\right) $ and $%
\mathrm{III}$ is the out region $t\in \left[ t_{\mathrm{out}},+\infty
\right) $ and $t_{\mathrm{out}}$, $t_{\mathrm{in}}$ are constants, $t_{%
\mathrm{out}}-t_{\mathrm{in}}=T$. We choose $t_{\mathrm{out}}=-t_{\mathrm{in}%
}=T/2$. The functions $\ _{-}\varphi _{n}\left( t\right) $ and $\
^{+}\varphi _{n}\left( t\right) $ and the distribution $N_{n}^{\mathrm{cr}}$
are found in Ref.~\cite{GavG96a}. The $T$-constant field can be considered
as slowly varying if
\begin{equation}
\sqrt{eE_{0}}T\gg \max \left( 1,m^{2}/eE_{0}\right) .  \label{T-large}
\end{equation}%
In this case, the leading contribution to the total number of pairs\ created
is formed in the range of $\left\vert p_{x}\right\vert <eE_{0}T/2$ and small
$\pi _{\perp }\ll eE_{0}T/2$ and has a form ~(\ref{asy4}).

(iii) A peak electric field $E\left( t\right) $ is composed of two parts.
One of them is increasing exponentially on the time interval $\mathrm{I}%
=\left( -\infty ,0\right] $, and reaches its maximal magnitude $E_{0}>0$ at
the end of the interval $t=0$. The second part decreases exponentially on
the time interval $\mathrm{II}=\left( 0,+\infty \right) $ having at $t=0$
the same magnitude $E_{0}.$ The vector potential $A_{x}\left( t\right) $ and
the field $E_{x}\left( t\right) $ are
\begin{equation}
E\left( t\right) =E_{0}\left\{
\begin{array}{l}
e^{k_{1}t}\,,\ \ t\in \mathrm{I}\, \\
e^{-k_{2}t}\,,\ \ t\in \mathrm{II}%
\end{array}%
\right. ,\ A_{x}\left( t\right) =E_{0}\left\{
\begin{array}{l}
k_{1}^{-1}\left( -e^{k_{1}t}+1\right) ,\ \ t\in \mathrm{I}\, \\
k_{2}^{-1}\left( e^{-k_{2}t}-1\right) \,,\ \ t\in \mathrm{II}%
\end{array}%
\right. ,  \label{ns4.0}
\end{equation}%
where $k_{1}$ and $k_{2}$ are positive constants. The functions $\ _{\zeta
}\varphi _{n}\left( t\right) $ and $\ ^{\zeta }\varphi _{n}\left( t\right) $
and the distribution $N_{n}^{\mathrm{cr}}$ are found in Ref.~ \cite%
{AdoGavGit16}. In particular, in the intervals $\mathrm{I}$ and $\mathrm{II}$
we have\textrm{\ }the following behavior%
\begin{eqnarray}
\ _{+}\varphi _{n}\left( t\right) &=&\;_{+}\mathcal{N}\exp \left( i\pi \nu
_{1}/2\right) y_{2}^{1}\left( \eta _{1}\right) \,,\,\ _{-}\varphi _{n}\left(
t\right) =\;_{-}\mathcal{N}\exp \left( -i\pi \nu _{1}/2\right)
y_{1}^{1}\left( \eta _{1}\right) \,,\ \ t\in \mathrm{I}\,;  \notag \\
\ ^{+}\varphi _{n}\left( t\right) &=&\;^{+}\mathcal{N}\exp \left( -i\pi \nu
_{2}/2\right) y_{1}^{2}\left( \eta _{2}\right) \,,\,\ ^{-}\varphi _{n}\left(
t\right) =\;^{-}\mathcal{N}\exp \left( i\pi \nu _{2}/2\right)
y_{2}^{2}\left( \eta _{2}\right) \,,\ \ t\in \mathrm{II}\,.  \notag \\
y_{1}^{j}\left( \eta _{j}\right) &=&e^{-\eta _{j}/2}\eta _{j}^{\nu _{j}}\Phi
\left( a_{j},c_{j};\eta _{j}\right) \,,\,\ y_{2}^{j}\left( \eta _{j}\right)
=e^{\eta _{j}/2}\eta _{j}^{-\nu _{j}}\Phi \left( 1-a_{j},2-c_{j};-\eta
_{j}\right) \,,  \label{i.4.1}
\end{eqnarray}%
where $\Phi \left( a,c;\eta \right) $ is confluent hypergeometric function
\cite{BatE53} and
\begin{eqnarray*}
\eta _{1} &=&ih_{1}e^{k_{1}t}\,,\ \ \eta _{2}=ih_{2}e^{-k_{2}t}\,,\ \
h_{j}=2eE_{0}k_{j}^{-2},\ \ j=1,2\,, \\
c_{j} &=&1+2\nu _{j}\,,\ \ a_{j}=\frac{1}{2}\left( 1+\chi \right) +\left(
-1\right) ^{j}\frac{i\pi _{j}}{k_{j}}+\nu _{j}\,, \\
\nu _{j} &=&\frac{i\omega _{j}}{k_{j}},\ \ \omega _{j}=\sqrt{\pi
_{j}^{2}+\pi _{\perp }^{2}},\ \ \pi _{j}=p_{x}-\left( -1\right) ^{j}\frac{%
eE_{0}}{k_{j}}\,.
\end{eqnarray*}%
The slowly varying peak field corresponds to small values of $k_{1}$ and $%
k_{2}$, and is characterized by the following criterion,
\begin{equation}
\min \left( h_{1},h_{2}\right) \gg \max \left( 1,m^{2}/eE_{0}\right) .\,
\label{4.1}
\end{equation}%
In this case, the main contributions to $N_{n}^{\mathrm{cr}}$ are formed in
the ranges $\pi _{\bot }<\pi _{1}\leqslant eE_{0}/k_{1}$ and $%
-eE_{0}/k_{2}<\pi _{2}<-\pi _{\bot }$, where they have the following forms:%
\begin{eqnarray}
N_{n}^{\mathrm{cr}} &\approx &\exp \left[ -\frac{2\pi }{k_{1}}\left( \omega
_{1}-\pi _{1}\right) \right] ,\mathrm{\;}\pi _{\bot }<\pi _{1}\leqslant
eE_{0}/k_{1},  \notag \\
N_{n}^{\mathrm{cr}} &\approx &\exp \left[ -\frac{2\pi }{k_{2}}\left( \omega
_{2}+\pi _{2}\right) \right] ,\mathrm{\;}-eE_{0}/k_{2}<\pi _{2}<-\pi _{\bot
}.  \label{4.10}
\end{eqnarray}

In the examples under discussion, the switch-on and switch-off regimes are
described by nearly the same functional form; that is, increasing and
decreasing components of the fields are almost symmetric. We have an
essentially asymmetric configuration in the case of the peak field, when the
field switches abruptly on at $t=0$; that is, $k_{1}$ is sufficiently large,%
\begin{equation}
eE_{0}k_{1}^{-2}\ll 1,\;\;\omega _{1}/k_{1}\ll 1,  \label{5.6}
\end{equation}%
while the parameter $k_{2}>0$ is arbitrary and includes the case of a smooth
switching off. We refer to this configuration as the exponentially decaying
electric field, see Ref.~\cite{AdoGavGit16} for details. The case of slowly
varying field we have when%
\begin{equation*}
h_{2}\gg \max \left( 1,m^{2}/eE_{0}\right) .
\end{equation*}%
In this case, the leading contribution to the total number of pairs\ created
from vacuum is formed in the range $-eE_{0}/k_{2}<\pi _{2}<-\pi _{\bot }$.
In this range, $N_{n}^{\mathrm{cr}}$ coincides with the form given by the
second line in Eq.(\ref{4.10}). Note that due to the invariance of the mean
numbers $N_{n}^{\mathrm{cr}}$ under the simultaneous change $%
k_{1}\leftrightarrows k_{2}$ and $\pi _{1}\leftrightarrows -\pi _{2}$, one
can easily transform this situation to the case with a large $k_{2}$ and
arbitrary $k_{1}>0$.

As it follows from calculations in the exactly solvable cases, for slowly
varying electric fields differential mean numbers of electron-positron pairs
created from the vacuum\ $N_{n}^{\mathrm{cr}}$ are quasiconstant over the
wide range of the longitudinal momentum $p_{x}$\ for any given transversal
momenta, although these distributions $N_{n}^{\mathrm{cr}}$ are different
for different field configurations. Furthermore, in all these cases, there
exist wide subranges, in which these distributions \ $N_{n}^{\mathrm{cr}}$
coincide with the corresponding distributions $N_{n}^{0}$ in a constant
electric field, given by Eq.~(\ref{asy4}). We call this phenomenon a
stabilization of the particle creation effect. In these subranges the mean
numbers $N_{n}^{\mathrm{cr}}$ hardly depend\textrm{\ }of the details of
switching on and off of electric field.

The total number of pairs\ created from a vacuum by a uniform electric field
is proportional to the space volume $V_{\left( d-1\right) }$ as $N^{\mathrm{%
cr}}=V_{\left( d-1\right) }n^{\mathrm{cr}}$, where $d$ labels the space-time
dimensions, and the corresponding densities $n^{\mathrm{cr}}$ have the form
\begin{equation}
n^{\mathrm{cr}}=\frac{J_{(d)}}{(2\pi )^{d-1}}\int d\mathbf{p}N_{n}^{\mathrm{%
cr}}\,.  \label{asy5}
\end{equation}%
In deriving Eq. (\ref{asy5}) a sum over all momenta $\mathbf{p}$\ was
transformed into an integral and summation over spin projections was
fulfilled, $J_{(d)}=2^{\left[ d/2\right] -1}$.\emph{\ }In slowly varying
fields, the total increment of the longitudinal kinetic momentum, which is $%
\Delta U=e\left\vert A_{x}\left( +\infty \right) -A_{x}\left( -\infty
\right) \right\vert $, is large and can be used as a large parameter.\emph{\
}Then the integral in the right-hand side of Eq. (\ref{asy5}) can be
approximated by an integral over a subrange $\Omega $\ that gives the
dominant contribution with respect to the total increment to the mean number
of created particles,%
\begin{equation}
\Omega :n^{\mathrm{cr}}\approx \tilde{n}^{\mathrm{cr}}=\frac{J_{(d)}}{(2\pi
)^{d-1}}\int_{\mathbf{p\in }\Omega }d\mathbf{p}N_{n}^{\mathrm{cr}}\,.
\label{asy6}
\end{equation}%
The dominant contributions $\tilde{n}^{\mathrm{cr}}$\ are proportional to
increments of the longitudinal kinetic momentum, which, in general, differ
for different fields and, for example, have the following forms in the
exactly solvable cases (i), (ii), and (iii):%
\begin{eqnarray}
\mathrm{(i)}\;\Delta U_{\mathrm{S}} &=&2eE_{0}T_{\mathrm{S}}\;\mathrm{%
for\;Sauter}\text{\textrm{-}}\mathrm{like\ field},  \notag \\
\mathrm{(ii)}\;\Delta U_{\mathrm{T}} &=&eE_{0}T\;\;\mathrm{for\;}T\text{%
\textrm{-}}\mathrm{const.\ field},  \notag \\
\mathrm{(iii)}\;\Delta U_{\mathrm{p}} &=&eE_{0}\left(
k_{1}^{-1}+k_{2}^{-1}\right) \;\mathrm{for\;a\ peak\ field}.  \label{4.21}
\end{eqnarray}%
We note that $\Delta U_{\mathrm{p}}$ in Eq.~(\ref{4.21}) corresponds to the
case of an exponentially decaying field at $k_{1}^{-1}\rightarrow 0$.

In terms of the introduced quantities (\ref{4.21}), the densities $\tilde{n}%
^{\mathrm{cr}}$ in the exactly solvable cases under consideration have the
following forms\footnote{%
Note that the derivation of total quantities for the Sauter-like case in
Refs. \cite{GavG96a} is given for $\lambda _{0}>1$. However, the final form
of $\delta =\sqrt{\pi }\Psi \left( \frac{1}{2},\frac{2-d}{2};\pi \frac{m^{2}%
}{eE_{0}}\right) $ is given correctly for arbitrary $m^{2}/eE_{0}$.} \cite%
{GavG96a,AdoGavGit16} (see \cite{AdoGavGit15} for more details\textrm{\ }):
\begin{eqnarray}
\mathrm{(i)}\;\tilde{n}^{\mathrm{cr}} &=&r^{\mathrm{cr}}\frac{\Delta U_{%
\mathrm{S}}}{2eE_{0}}\delta ,\;  \notag \\
\mathrm{(ii)}\;\tilde{n}^{\mathrm{cr}} &=&r^{\mathrm{cr}}\frac{\Delta U_{%
\mathrm{T}}}{eE_{0}},  \notag \\
\mathrm{(iii)}\;\tilde{n}^{\mathrm{cr}} &=&r^{\mathrm{cr}}\frac{\Delta U_{%
\mathrm{p}}}{eE_{0}}G\left( \frac{d}{2},\pi \frac{m^{2}}{eE_{0}}\right) .
\label{4.23}
\end{eqnarray}%
where%
\begin{eqnarray}
&&r^{\mathrm{cr}}=\frac{J_{(d)}\left( eE_{0}\right) ^{d/2}}{(2\pi )^{d-1}}%
\exp \left\{ -\pi \frac{m^{2}}{eE_{0}}\right\} ,\ G\left( \alpha ,x\right)
=\int_{1}^{\infty }\frac{ds}{s^{\alpha +1}}e^{-x\left( s-1\right)
}=e^{x}x^{\alpha }\Gamma \left( -\alpha ,x\right) ,  \notag \\
&&\delta =\int_{0}^{\infty }dtt^{-1/2}(t+1)^{-\left( d+1\right) /2}\exp
\left( -t\pi \frac{m^{2}}{eE_{0}}\right) =\sqrt{\pi }\Psi \left( \frac{1}{2},%
\frac{2-d}{2};\pi \frac{m^{2}}{eE_{0}}\right) \,.  \label{4.19a}
\end{eqnarray}%
Here $\Gamma \left( -\alpha ,x\right) $ is the incomplete gamma function and
$\Psi \left( a,b;x\right) $ is the confluent hypergeometric function \cite%
{BatE53}. Equating the densities $n^{\mathrm{cr}}$ for Sauter-like field (i)
and for the peak field (iii) to the density $n^{\mathrm{cr}}$ for the $T$%
-constant field (ii), we find an effective time $T_{eff}$ of the field
duration in both cases,%
\begin{eqnarray}
\mathrm{(i)}\;T_{eff} &=&T_{\mathrm{S}}\delta ,  \notag \\
\mathrm{(iii)}\;T_{eff} &=&\left( k_{1}^{-1}+k_{2}^{-1}\right) G\left( \frac{%
d}{2},\pi \frac{m^{2}}{eE_{0}}\right) .  \label{4.24}
\end{eqnarray}%
Note that the effective time $T_{eff}$ for an exponentially decaying field
is given by the second line in~Eq. (\ref{4.24}) as $k_{1}^{-1}\rightarrow 0$%
. By the definition $T_{eff}=T$ for the $T$-constant field. One can say that
the Sauter-like field, the peak electric field, and the exponentially
decaying field with the same $T_{eff}$ are equivalent to the $T$-constant
field with respect to the pair production. Note that the factors $G$ and $%
\delta $ in Eq.~(\ref{4.19a}) for a weak electric field ($m^{2}/eE_{0}\gg $\
$1$) \ and for a strong enough electric field ($m^{2}/eE_{0}\ll 1$) can be
approximated as
\begin{eqnarray}
G\left( \frac{d}{2},\pi \frac{m^{2}}{eE_{0}}\right) &\approx &\frac{eE_{0}}{%
\pi m^{2}},\ \ \delta \approx \frac{\sqrt{eE_{0}}}{m}\ ,\mathrm{\;}\frac{%
m^{2}}{eE_{0}}\gg 1;  \notag \\
G\left( \frac{d}{2},\pi \frac{m^{2}}{eE_{0}}\right) &\approx &\frac{2}{d},\
\ \delta \approx \frac{\sqrt{\pi }\Gamma \left( d/2\right) }{\Gamma \left(
d/2+1/2\right) },\mathrm{\;}m^{2}/eE_{0}\ll 1.  \label{4.26}
\end{eqnarray}

Additionally, we note that one can compare the time scales for the cases
(i), (ii), and (iii), given by Eqs. (\ref{asy1}), (\ref{T-large}) and (\ref%
{4.1}), for the same $E_{0}$ and kinetic momentum increments, $\Delta U_{%
\mathrm{S}}=\Delta U_{\mathrm{T}}=\Delta U_{\mathrm{p}}$ (in this case $2T_{%
\mathrm{S}}=T=k_{1}^{-1}+k_{2}^{-1}$). One can see that the condition (\ref%
{T-large}){\Large \ }is stronger than that given by Eqs. (\ref{asy1}) and (%
\ref{4.1}) if fields are weak, whereas they are equivalent if fields are
strong. For this reason, defining the scale $\Delta t_{\mathrm{st}}^{\mathrm{%
m}}$ in general terms, we choose the form (\ref{svf2}).

Let us turn to the vacuum-to-vacuum transition probability $P_{v}$ defined
by Eq.~(\ref{vacprob}) in the Appendix.\textrm{\ }It is given by similar
forms for the Sauter-like field (i), the $T$-constant field (ii), and the
peak field (iii), respectively, with the corresponding $N^{\mathrm{cr}}$
\cite{GavG96a,AdoGavGit16}:%
\begin{eqnarray}
&&P_{v}=\exp \left( -\mu N^{\mathrm{cr}}\right) ,\;\;\mu =\sum_{l=0}^{\infty
}\frac{\epsilon _{l+1}}{(l+1)^{d/2}}\exp \left( -l\pi \frac{m^{2}}{eE_{0}}%
\right) ,  \notag \\
\mathrm{(i)}\;\epsilon _{l} &=&\epsilon _{l}^{\mathrm{S}}=\delta ^{-1}\sqrt{%
\pi }\Psi \left( \frac{1}{2},\frac{2-d}{2};l\pi \frac{m^{2}}{eE_{0}}\right) ,
\notag \\
\mathrm{(ii)}\; &&\epsilon _{l}=\epsilon _{l}^{\mathrm{T}}=1,  \notag \\
\mathrm{(iii)}\;\epsilon _{l} &=&G\left( \frac{d}{2},l\pi \frac{m^{2}}{eE_{0}%
}\right) \left[ G\left( \frac{d}{2},\pi \frac{m^{2}}{eE_{0}}\right) \right]
^{-1}.  \label{4.29}
\end{eqnarray}

In the case of a weak field{\large \ (}$m^{2}/eE_{0}>1${\large ),}$%
\;\epsilon _{l}^{\mathrm{S}}\approx l^{-1/2}${\large \ }for the Sauter-like
field{\large , }$\epsilon _{l}\approx l^{-1}${\large \ }for the peak field,
and{\large \ }$\exp \left( -\pi m^{2}/eE_{0}\right) \ll 1${\large . }Then $%
\mu \approx 1$\ for all the cases in Eq.~(\ref{4.29}) and we have a
universal relation{\large \ }$N^{\mathrm{cr}}\approx \ln P_{v}^{-1}$. In the
case of a strong field ($m^{2}/eE_{0}\ll 1$), all the terms with different $%
\epsilon _{l}^{\mathrm{S}}${\large \ }and $\epsilon _{l}$\ contribute
significantly to the sum in Eq.~(\ref{4.29}) if $l\pi m^{2}/eE_{0}\lesssim 1$%
, and the quantities $\mu $\ for the Sauter-like and peak\ fields differ
essentially from the case of a $T$-constant field. Consequently, in this
situation, one cannot derive a universal relation between $N^{\mathrm{cr}}$
and $P_{v}$\ from particular cases given by Eq.~(\ref{4.29}). In addition,
it should be noted that in the case of a strong field, when known
semiclassical approaches are{\Large \ }not applicable, the probability $%
P_{v} $\ (unlike the total number {\large \ }$N^{\mathrm{cr}}$) no longer
has a direct relation to vacuum mean values of the physical quantities
discussed above. Therefore, to study a universal behavior of the vacuum
instability in slowly varying strong electric fields one should derive first
a universal form for the total density{\large \ }$\tilde{n}^{\mathrm{cr}}$%
{\large .}

\section{Universal behavior of the vacuum instability in slowly varying
strong electric fields \label{S3}}

\subsection{Total density of created pairs}

If the electric field is not very strong, mean numbers $N_{n}^{\mathrm{cr}}$
of created pairs (or distributions) at the final time instant are
exponentially small, $N_{n}^{\mathrm{cr}}\ll 1$. In this case the
probability of the vacuum to remain a vacuum and probabilities of particle
scattering and pair creation have simple representations in terms of these
numbers,%
\begin{equation}
\left\vert w_{n}\left( +-|0\right) \right\vert ^{2}\approx N_{n}^{\mathrm{cr}%
},\;\left\vert w_{n}\left( -|-\right) \right\vert ^{2}\approx \left(
1+N_{n}^{\mathrm{cr}}\right) ,\;P_{v}\approx 1-\sum_{n}N_{n}^{\mathrm{cr}}.
\label{a3.1}
\end{equation}%
The latter relations are often used in semiclassical calculations\emph{\ }to
find $N_{n}^{\mathrm{cr}}$\ and the total number of created pairs $N^{%
\mathrm{cr}}=\sum_{n}N_{n}^{\mathrm{cr}}$ from the representation of $P_{v}$%
\ given by Schwinger's effective action.

However, when the electric field cannot be considered as a weak one (e.g. in
some situations in astrophysics and condensed matter), the mean numbers $%
N_{n}^{\mathrm{cr}}$ can achieve their limited values $N_{n}^{\mathrm{cr}%
}\rightarrow 1$ already at finite time instants $t$ and the sum $N^{\mathrm{%
cr}}$ cannot be considered as a small quantity. Moreover, for slowly varying
strong electric fields this sum is proportional to the large parameter $%
T_{eff}/\Delta t_{\mathrm{st}}$. In such a case relations (\ref{a3.1}) are
not correct anymore.{\large \ }However, as shown next, for arbitrarily
slowly varying strong electric field one can derive in the leading-term
approximation an universal form for the total density of created pairs.

Let us define the range $D\left( t\right) $ as follows:
\begin{equation}
D\left( t\right) :\left\langle P_{x}\left( t\right) \right\rangle
<0,\;\;\left\vert \left\langle P_{x}\left( t\right) \right\rangle
\right\vert \gg \pi _{\perp }.  \label{4.31a}
\end{equation}%
In this range the longitudinal kinetic momentum $\left\langle P_{x}\left(
t\right) \right\rangle =p_{x}-U\left( t\right) $ is negative and big enough.
If $p_{x}$ components of the particle momentum belong to the range $D\left(
t\right) $, then the particle energy is primarily determined by an increment
of the longitudinal kinetic momentum\emph{, }$U\left( t\right) -U\left( t_{%
\mathrm{in}}\right) $,\emph{\ }during the time interval\emph{\ }$t-t_{%
\mathrm{in}}$ and $\left\langle P_{x}\left( t\right) \right\rangle =$ $%
\left\langle P_{x}\left( t_{\mathrm{in}}\right) \right\rangle -\left[
U\left( t\right) -U\left( t_{\mathrm{in}}\right) \right] $. Note that $%
D\left( t\right) \subset D\left( t^{\prime }\right) $ if $t<t^{\prime }$.
The leading term of the total number density of created pairs, $\tilde{n}^{%
\mathrm{cr}}$, is formed over the range $D\left( t_{\mathrm{out}}\right) $,
that is, the range $D\left( t_{\mathrm{out}}\right) $ is chosen as a
realization of the subrange $\Omega $\ in Eq.~(\ref{asy6}).

In the case when the electric field does not switch abruptly on and off,
that is, the field slowly weakens at $t\rightarrow \pm \infty $ and one of
the time instants $t_{\mathrm{in}}$ and $t_{\mathrm{out}}$ , or both are
infinite $t_{\mathrm{in}}\rightarrow -\infty $ and $t_{\mathrm{out}%
}\rightarrow \infty $, one can ignore exponentially small contributions to $%
\tilde{n}^{\mathrm{cr}}$ from the time intervals $\left( t_{\mathrm{in}},t_{%
\mathrm{in}}^{eff}\right] $ and $\left( t_{\mathrm{out}}^{eff},t_{\mathrm{out%
}}\right) $, where electric fields are much less than the maximum field $%
E_{0}$, $E\left( t_{\mathrm{in}}^{eff}\right) ,E\left( t_{\mathrm{out}%
}^{eff}\right) \ll E_{0}$. Thus, in the general case it is enough to
consider a finite interval $\left( t_{\mathrm{in}}^{eff},t_{\mathrm{out}%
}^{eff}\right] $. Denoting $t_{1}=t_{\mathrm{in}}^{eff}$ and $t_{M+1}=t_{%
\mathrm{out}}^{eff}$, we divide this interval into $M$ intervals $\Delta
t_{i}=t_{i+1}-t_{i}>0$, $i=1,...,M$, $\sum_{i=1}^{M}\Delta t_{i}=t_{\mathrm{%
out}}^{eff}-t_{\mathrm{in}}^{eff}$. We suppose that Eqs.~(\ref{svf1}) and (%
\ref{svf2}) hold true for all the intervals, respectively. That allows us to
treat the electric field as approximately constant within each interval, $%
\overline{E(t)}\approx \overline{E}(t_{i})$, for $t\in \left( t_{i},t_{i+1}%
\right] $. Note that inside of each interval $\Delta t_{i}$ abrupt changes
of the electric field $E(t)$, whose duration is much less than $\Delta t_{i}$%
, cannot change significantly the total value of $\tilde{n}^{\mathrm{cr}}$,
since $N_{n}^{\mathrm{cr}}\leq 1$ for fermions. Using Eqs.~(\ref{4.21}) and (%
\ref{4.23}) for the case of $T$-constant field, we can represent $\tilde{n}^{%
\mathrm{cr}}$ as the following sum
\begin{eqnarray}
&&\tilde{n}^{\mathrm{cr}}=\sum_{i=1}^{M}\Delta \tilde{n}_{i}^{\mathrm{cr}%
},\;\Delta \tilde{n}_{i}^{\mathrm{cr}}\approx \frac{J_{(d)}}{(2\pi )^{d-1}}%
\int_{e\overline{E}(t_{i})}^{e\overline{E}(t_{i})\left( t_{i}+\Delta
t_{i}\right) }dp_{x}\int_{\sqrt{\lambda _{i}}<K_{\bot }}d\mathbf{p}_{\bot
}N_{n}^{\left( i\right) }\,,  \notag \\
&&N_{n}^{\left( i\right) }=e^{-\pi \lambda _{i}},\ \ \lambda _{i}=\frac{\pi
_{\bot }^{2}}{e\overline{E}(t_{i})}\,,\   \label{uni1}
\end{eqnarray}%
where $K_{\bot }$ is any given number satisfying the condition$\;\sqrt{e%
\overline{E}(t_{i})}\Delta t_{i}\gg K_{\bot }^{2}\gg \max \left\{ 1,m^{2}/e%
\overline{E}(t_{i})\right\} $. Taking into account Eq.~(\ref{4.31a}), we
represent the variable $p_{x}$ as follows
\begin{equation}
p_{x}=U\left( t\right) ,\!\!\;\;U\left( t\right) =\int_{t_{\mathrm{in}%
}}^{t}dt^{\prime }eE\left( t^{\prime }\right) +U\left( t_{\mathrm{in}%
}\right) .  \label{uni2b}
\end{equation}%
Then neglecting small contributions to the integral (\ref{uni1}), we find
the following universal form for the total density of created pairs in the
leading-term approximation for a slowly varying, but otherwise arbitrary
strong electric field%
\begin{equation}
\tilde{n}^{\mathrm{cr}}\approx \frac{J_{(d)}}{(2\pi )^{d-1}}\int_{t_{\mathrm{%
in}}}^{t_{\mathrm{out}}}dteE\left( t\right) \int d\mathbf{p}_{\bot }N_{n}^{%
\mathrm{uni}},\ \ N_{n}^{\mathrm{uni}}=\exp \left[ -\pi \frac{\pi _{\bot
}^{2}}{eE\left( t\right) }\right] .  \label{uni2}
\end{equation}%
Note that $N_{n}^{\mathrm{uni}}$ is written in a universal form which can be
used to calculate any total characteristics of the pair creation effect.\
After the integration over $\mathbf{p}_{\bot }$, we finally obtain%
\begin{equation}
\tilde{n}^{\mathrm{cr}}=\frac{J_{(d)}}{(2\pi )^{d-1}}\int_{t_{\mathrm{in}%
}}^{t_{\mathrm{out}}}dt\left[ eE\left( t\right) \right] ^{d/2}\exp \left\{
-\pi \frac{m^{2}}{eE\left( t\right) }\right\} .  \label{uni3}
\end{equation}

These universal forms can be derived for bosons as well, if we are
restricting them to forms of external electric fields, namely, fields that
have no abrupt variations of $E(t)$ that can produce significant growth of $%
N_{n}^{\mathrm{cr}}$ on a finite time interval. In fact, in this case we
have to include in the range $D\left( t\right) $ the only subranges where $%
N_{n}^{\mathrm{cr}}\leq 1$. In this case the universal forms for bosons are
the same (\ref{uni2}) and (\ref{uni3}) assuming that $J_{(d)}$ is the number
of the boson spin degrees of freedom, in particular, $J_{(d)}=1$ for scalar
particles and $J_{(4)}=3$ for vector particles.

Using the identity $-\ln \left( 1-N_{n}^{\mathrm{uni}}\right) =N_{n}^{%
\mathrm{uni}}+\left( N_{n}^{\mathrm{uni}}\right) ^{2}\ldots $, in the same
manner one can derive a universal form of the vacuum-to-vacuum transition
probability $P_{v}$ defined for fermions by Eq.~(\ref{vacprob}) in the
Appendix. First, we write%
\begin{equation}
P_{v}\approx \exp \left[ -\frac{V_{\left( d-1\right) }J_{(d)}}{(2\pi )^{d-1}}%
\sum_{l=1}^{\infty }\int_{t_{\mathrm{in}}}^{t_{\mathrm{out}}}dteE\left(
t\right) \int d\mathbf{p}_{\bot }\left( N_{n}^{\mathrm{uni}}\right) ^{l}%
\right] .  \label{uni4}
\end{equation}%
Then, performing the integration over $\mathbf{p}_{\bot }$, we obtain that
for\textrm{\ }fermions this universal form reads%
\begin{equation}
P_{v}\approx \exp \left\{ -\frac{V_{\left( d-1\right) }J_{(d)}}{(2\pi )^{d-1}%
}\sum_{l=1}^{\infty }\int_{t_{\mathrm{in}}}^{t_{\mathrm{out}}}dt\frac{\left[
eE\left( t\right) \right] ^{d/2}}{l^{d/2}}\exp \left[ -\pi \frac{lm^{2}}{%
eE\left( t\right) }\right] \right\} .  \label{uni5}
\end{equation}

Taking into account that universal forms of $\tilde{n}^{\mathrm{cr}}$ for
bosons are given by formulas similar to Eqs.~(\ref{uni2}) and (\ref{uni3})
and using the definition of the vacuum-to-vacuum transition probability $%
P_{v}^{\left( boson\right) }$ for bosons obtained in Refs. \cite{FGS,GavGT06}%
,
\begin{equation}
P_{v}^{\left( boson\right) }=\exp \left[ -\sum_{n}\ln \left( 1+N_{n}^{%
\mathrm{cr}}\right) \right] ,  \label{uni6a}
\end{equation}%
we finally get in the Bose case the following universal form%
\begin{equation}
P_{v}^{\left( boson\right) }\approx \exp \left\{ -\frac{V_{\left( d-1\right)
}J_{(d)}}{(2\pi )^{d-1}}\sum_{l=1}^{\infty }\int_{t_{\mathrm{in}}}^{t_{%
\mathrm{out}}}dt\left( -1\right) ^{l-1}\frac{\left[ eE\left( t\right) \right]
^{d/2}}{l^{d/2}}\exp \left[ -\pi \frac{lm^{2}}{eE\left( t\right) }\right]
\right\} ,  \label{uni6}
\end{equation}%
where $J_{(d)}$ is the number of boson spin degrees of freedom.

Using Eqs.~(\ref{uni3}) and (\ref{uni5}), one obtains precisely expressions (%
\ref{4.23}) and (\ref{4.29}) that are found for the total densities and the
vacuum-to-vacuum transition probabilities when directly adopting the slowly
varying field approximation to the exactly solvable cases. Comparing Eqs.~(%
\ref{uni3}) and (\ref{uni6}) with the exact results obtained for bosons \cite%
{GavG96a,AdoGavGit16}, one finds precise agreement too. Thus, we have an
independent confirmation of the universal forms obtained above.

One can see that the obtained universal forms have specially simple forms in
two limited cases, for a weak electric field ($m^{2}/eE_{0}\gg $\ $1$), when
the term $\left[ eE\left( t\right) \right] ^{d/2}$ can be approximated by
its maximal value $\left[ eE_{0}\right] ^{d/2}$, and for a strong enough
electric field ($m^{2}/eE_{0}\ll 1$), when there exist time intervals where $%
m^{2}/eE\left( t\right) \ll 1$ and approximations of the type
\begin{equation}
\exp \left[ -\frac{\pi lm^{2}}{eE\left( t\right) }\right] =1-\frac{\pi lm^{2}%
}{eE\left( t\right) }+\ldots \;  \label{strong}
\end{equation}%
are available. Consider, for example, the case of a strong Gauss pulse,
\begin{equation}
E\left( t\right) =E_{0}\exp \left[ -\left( t/T_{G}\right) ^{2}\right] ,
\label{G1}
\end{equation}%
with a large parameter $T_{G}\rightarrow \infty $. In this case we do not
have an exact solution of the Dirac equation and known semiclassical
approximations are not applicable. However, using approximation (\ref{strong}%
), we find from Eqs.~(\ref{uni3}) and (\ref{uni5}) the leading terms as
\begin{equation}
\tilde{n}^{\mathrm{cr}}\approx \frac{J_{(d)}\left( eE_{0}\right) ^{d/2}T_{G}%
}{d(2\pi )^{d-2}},\ \ P_{v}\approx \exp \left[ -V_{\left( d-1\right) }\tilde{%
n}^{\mathrm{cr}}\sum_{l=1}^{\infty }l^{-d/2}\right] .  \label{G2}
\end{equation}

The representations (\ref{uni5}) and (\ref{uni6}) coincide with the leading
term approximation of derivative expansion results from field-theoretic
calculations obtained in Refs.~\cite{DunnH98,GusSh99} for $d=3$ and $d=4$%
{\large .} In this approximation the probability $P_{v}$ was derived from a
formal expansion in increasing numbers of derivatives of the background
field strength for Schwinger's effective action:%
\begin{equation}
S=S^{\left( 0\right) }[F_{\mu \nu }]+S^{\left( 2\right) }[F_{\mu \nu
},\partial _{\mu }F_{\nu \rho }]+...  \label{uni7b}
\end{equation}%
where $S^{\left( 0\right) }$ involves no derivatives of the background field
strength $F_{\mu \nu }$ (that is, $S^{\left( 0\right) }$ is a locally
constant field approximation for $S$), while the first correction $S^{\left(
2\right) }$ involves two derivatives of the field strength, and so on, see
Ref.~\cite{Dunn04} for a review. In fact, it is the possibility to adopt a
locally constant field approximation which makes the effect universal.%
{\large \ }"

It was found that%
\begin{equation}
P_{v}=\exp \left( -2\mathrm{Im}S^{\left( 0\right) }\right) .  \label{uni7}
\end{equation}%
In the derivative expansion the fields are assumed to vary very slowly and
satisfy the condition (\ref{svf1}). A very convenient formalism for doing
such an expansion is the worldline formalism, see \cite{Schub01} for the
review, in which the effective action is written as a quantum-mechanical
path integral.

However, for a general background field, it is extremely difficult to
estimate and compare the magnitude of various terms in the derivative
expansion. Only under the assumption $m^{2}/eE_{0}>1$ can one demonstrate
that the derivative expansion is completely consistent with the
semiclassical WKB analysis of the imaginary part of the effective action
\cite{DunnH99}. It is shown only for a constant electric field that Eq.~(\ref%
{uni7}) is given exactly by the semiclassical WKB limit when the leading
order of fluctuations is taken into account \cite{GordonS15}.

It should be stressed that unlike the authors of Refs.~\cite{DunnH98,GusSh99}%
, we derive Eqs.~(\ref{uni5}) and (\ref{uni6}) in the framework of the
general exact formulation of strong-field QED \cite{FGS,GavGT06}, where $%
P_{v}$ are defined by Eqs.~(\ref{vacprob}){\large \ }and (\ref{uni6a}),
respectively. Therefore we obtain Eqs.~(\ref{uni5}) and (\ref{uni6})
independently from the derivative expansion approach and the obtained result
holds true for any strong field under consideration.{\large \ }Thus, it is
proven that Eq.~(\ref{uni7}) is given exactly by the semiclassical WKB limit
for arbitrarily slowly varying electric field.

\subsection{Time evolution of vacuum instability}

In this section details of the time evolution of vacuum instability effects
are of interest. In particular, the study of the time evolution of the mean
electric current, energy, and momentum\emph{\ }provides us with new
characteristics of the effect, related, in particular, with the
backreaction. Due to the translational invariance of the spatially uniform
external field, all the corresponding mean values are proportional to the
space volume. Therefore, it is enough to calculate the vacuum mean values of
the current density vector $\langle j^{\mu }(t)\rangle $ and of the
energy-momentum tensor (EMT) $\langle T_{\mu \nu }(t)\rangle $, defined by
Eq.~(\ref{int1}); see the Appendix. Note that these densities depend on the
initial vacuum, on the evolution of the electric field from the initial time
instant\ up to the current time instant $t$, but they do not depend on the
further history of the system and definition of particle-antiparticle at the
time $t$.

Let us consider the time dependence of the current density vector $\langle
j^{\mu }(t)\rangle $\ and of the EMT $\langle T_{\mu \nu }(t)\rangle $,
given by Eqs.~(\ref{A1.4}). Due to the uniform character of the
distributions $N_{n}^{\mathrm{cr}}$, only the diagonal matrix elements of
EMT differ from zero, in particular, for $d\neq 3$ only the longitudinal
current components are not zero. In $d=3$ dimensions, there are two
nonequivalent representations for $\gamma $-matrices, $\gamma ^{0}=\sigma
^{3},$ $\gamma ^{1}=i\sigma ^{2},$ $\gamma ^{2}=-i\left( \pm 1\right) \sigma
^{1},$ where $\sigma ^{i}$ are Pauli matrices, and representations with the
sign $+$ or $-$ in the round brackets correspond to different fermion
species, the so-called $+$ and $-$ fermions, respectively. Due to this fact,
a nonzero current component $\langle j^{2}(t)\rangle $ can exist. This fact
is related to the so-called Chern-Simons term in the effective action \cite%
{NiemiSem83,redlich}; see details in Ref.~\cite{GavGitY12}. However, if
there are both fermion species in a model, as it takes place, for example,
in the Dirac model of the graphene, then $\langle j^{2}(t)\rangle =0$.

It follows from Eqs.~(\ref{3.25}) and (\ref{A1.4}) that the nonzero terms $%
\mathrm{Re}\langle j^{\mu }(t)\rangle ^{p}\,$ and $\mathrm{Re}\langle T_{\mu
\nu }(t)\rangle ^{p}$ appear due to the vacuum instability. These terms are
growing with time due to an increase of the number of states that are
occupied by created pairs. In any system of Fermi particles the mean value $%
\langle j^{2}(t)\rangle $ is finite.

As a consequence of Eq.~(\ref{4.31a}){\large , }we have
\begin{equation}
i\partial _{t}\;^{\pm }\varphi _{n}\left( t\right) \approx \pm \left\vert
\left\langle P_{x}\left( t\right) \right\rangle \right\vert \;^{\pm }\varphi
_{n}\left( t\right) ,  \label{4.31b}
\end{equation}%
which means that at the time $t$ we deal with an ultrarelativistic particle\
and its kinetic momentum $\left\langle P_{x}\left( t\right) \right\rangle $
can be considered as a large parameter. Considering the time dependence of
means $\mathrm{Re}\,\langle j^{1}(t)\rangle ^{p}$ and $\mathrm{Re}\,\langle
T_{\mu \mu }(t)\rangle ^{p}$, we suppose that the time difference $t-t_{%
\mathrm{in}}$ is big enough to satisfy Eq.~(\ref{4.31b}). Using the exact
relation Eq.~(\ref{t4.4}) to express solutions $_{\pm }\psi _{n}$ via $^{\pm
}\psi _{n}$, and neglecting strongly oscillating terms, we find that the
leading contribution to the function $S^{p}(x,x^{\prime })$ (defined by Eq.~(%
\ref{3.25})) at $t\sim t^{\prime }$\ can be represented by the following
expression
\begin{equation}
S^{p}(x,x^{\prime })\approx -i\sum_{n}N_{n}^{\mathrm{cr}}\left[ ^{+}{\psi }%
_{n}(x)^{+}{\bar{\psi}}_{n}(x^{\prime })-\,^{-}{\psi }_{n}(x)^{-}{\bar{\psi}}%
_{n}(x^{\prime })\right] \,.  \label{4.30}
\end{equation}%
It is clear that for any large enough difference{\large \ }$t-t_{\mathrm{in}%
} ${\large \ }the sum over momentum $p$\ in the right-hand side of Eq. (\ref%
{4.30}) can be approximated by a sum over the range{\large \ }$D\left( t_{%
\mathrm{out}}\right) ${\large \ }that gives the dominant contribution to the
mean number of created particles with respect to the total increment of the
longitudinal kinetic momentum. Moreover, taking into account Eqs.~(\ref%
{4.31a}) and (\ref{uni2b}), we see that{\large \ }$D\left( t\right) \subset
D\left( t^{\prime }\right) \subset D\left( t_{\mathrm{out}}\right) ${\large %
\ }if{\large \ }$t<t^{\prime }<t_{\mathrm{out}}${\large \ }and for a given
difference{\large \ }$t-t_{\mathrm{in}}$ the dominant contribution to the
right-hand side of Eq. (\ref{4.30}) is from a subrange{\large \ }$D\left(
t\right) \subset D\left( t_{\mathrm{out}}\right) ${\large .}

We recall that, according to Eq.~(\ref{t4.10}), one can choose the
corresponding \textrm{in} and \textrm{out} Dirac solutions either with $\chi
=+1$ or with $\chi =-1$. Using this possibility, we choose $\chi =+1$ for $%
^{+}{\psi }_{n}(x)$ and $\chi =-1$ for $^{-}{\psi }_{n}(x).$ With such a
choice, taking into account that $\mathbf{p\in }D\left( t\right) $, we
simplify essentially the matrix structure of the representation (\ref{4.30}%
). Thus, after a summation over spin polarizations $\sigma $, we obtain the
following result:%
\begin{equation}
S^{p}(x,x^{\prime })\approx (\gamma P+m)\Delta ^{p}(x,x^{\prime }),
\label{4.32}
\end{equation}%
where the function $\Delta ^{p}(x,x^{\prime })$ reads%
\begin{eqnarray*}
&&\Delta ^{p}(x,x^{\prime })=-i\sum_{\mathbf{p\in }D\left( t\right) }N_{n}^{%
\mathrm{cr}}\left\vert \left\langle P_{x}\left( t\right) \right\rangle
\right\vert \exp \left[ i\mathbf{p}\left( \mathbf{r-r}^{\prime }\right) %
\right] \\
&&\times \left\{ \left( 1+\gamma ^{0}\gamma ^{1}\right) \left. \left[
\;^{+}\varphi _{n}\left( t\right) \;^{+}\varphi _{n}^{\ast }\left( t^{\prime
}\right) \right] \right\vert _{\chi =+1}+\left( 1-\gamma ^{0}\gamma
^{1}\right) \left. \left[ \;^{-}\varphi _{n}\left( t\right) \;^{-}\varphi
_{n}^{\ast }\left( t^{\prime }\right) \right] \right\vert _{\chi
=-1}\right\} .
\end{eqnarray*}

Using Eq. (\ref{4.32}) in~Eq. (\ref{A1.4}) and transforming the sum over all
momenta\textbf{\ }$\mathbf{p}$ into an integral, we find{\large \ }the
following representations for the vacuum means of current density and EMT
components:
\begin{align}
& \langle j^{1}(t)\rangle ^{p}\approx 2e\frac{V_{\left( d-1\right) }J_{(d)}}{%
(2\pi )^{d-1}}\int_{\mathbf{p\in }D\left( t\right) }d\mathbf{p}N_{n}^{%
\mathrm{cr}}\rho \left( t\right) \left\vert \left\langle P_{x}\left(
t\right) \right\rangle \right\vert ;  \notag \\
& \langle T_{00}(t)\rangle ^{p}\approx \langle T_{11}(t)\rangle
^{p}\,\approx \frac{V_{\left( d-1\right) }J_{(d)}}{(2\pi )^{d-1}}\int_{%
\mathbf{p\in }D\left( t\right) }d\mathbf{p}N_{n}^{\mathrm{cr}}\rho \left(
t\right) \left\langle P_{x}\left( t\right) \right\rangle ^{2},  \notag \\
& \langle T_{ll}(t)\rangle ^{p}\,\approx \frac{V_{\left( d-1\right) }J_{(d)}%
}{(2\pi )^{d-1}}\int_{\mathbf{p\in }D\left( t\right) }d\mathbf{p}N_{n}^{%
\mathrm{cr}}\rho \left( t\right) p_{l}^{2},\;l=2,...,D,  \notag \\
& \rho \left( t\right) =2\left\vert \left\langle P_{x}\left( t\right)
\right\rangle \right\vert \left\{ \left. \left\vert \;^{+}\varphi _{n}\left(
t\right) \right\vert ^{2}\right\vert _{\chi =+1}+\left. \left\vert
\;^{-}\varphi _{n}\left( t\right) \right\vert ^{2}\right\vert _{\chi
=-1}\right\} ,  \label{4.33}
\end{align}%
where $D=d-1$.

One can verify, taking into account Eq.~(\ref{4.31b}), that the functions $\
^{\zeta }\varphi _{n}\left( t\right) $\ can be approximated by their
asymptotics (\ref{t4.1a}) in the range{\large \ }$D\left( t\right) $ if the
instant value of the longitudinal kinetic momentum differs slightly at the
time instant $t$ from its final value, such that%
\begin{equation}
\left\vert \left\langle P_{x}\left( t_{\mathrm{out}}\right) \right\rangle
-\left\langle P_{x}\left( t\right) \right\rangle \right\vert \ll \left\vert
\left\langle P_{x}\left( t\right) \right\rangle \right\vert .  \label{4.34}
\end{equation}%
\ In a sense this means that the time instant $t$ is close enough to the
final time instant, $t\rightarrow t_{\mathrm{out}}$. We find that%
\begin{equation}
\left. \rho \left( t\right) \right\vert _{t\rightarrow t_{\mathrm{out}}}=%
\left[ V_{\left( d-1\right) }\left\vert \left\langle P_{x}\left( t_{\mathrm{%
out}}\right) \right\rangle \right\vert \right] ^{-1}.  \label{4.35}
\end{equation}%
Then taking into account Eq.~(\ref{asy6}), we obtain from ~Eq.~(\ref{4.33})
that
\begin{equation}
\left. \langle j^{1}(t)\rangle ^{p}\right\vert _{t\rightarrow t_{\mathrm{out}%
}}\approx 2e\tilde{n}^{\mathrm{cr}},  \label{4.36a}
\end{equation}%
where{\large \ }$\tilde{n}^{\mathrm{cr}}${\large \ }is given by Eqs.~(\ref%
{uni2}) and (\ref{uni3}){\large .} It means that dominant contributions to
the mean numbers $N_{n}^{\mathrm{cr}}$ of created particles are formed
before the time instant $t$ that satisfies Eq.~(\ref{4.34})\emph{. }For $%
t>t_{\mathrm{out}}$, the pair production stops, vacuum polarization effects
disappear, and quantities ~(\ref{4.33}) for $t>t_{\mathrm{out}}$ maintain
their values at $t=t_{\mathrm{out}}$. Using Eq.~(\ref{4.35}), we obtain that
\begin{eqnarray}
&&\left. \langle j^{1}(t)\rangle \right\vert _{t>t_{\mathrm{out}}}\approx
\left. \langle j^{1}(t)\rangle ^{p}\right\vert _{t\rightarrow t_{\mathrm{out}%
}}\approx 2e\tilde{n}^{\mathrm{cr}},\;  \notag \\
&&\left. \langle T_{00}(t)\rangle \right\vert _{t>t_{\mathrm{out}}}\approx
\left. \langle T_{11}(t)\rangle \right\vert _{t>t_{\mathrm{out}}}\approx
\left. \langle T_{11}(t)\rangle ^{p}\right\vert _{t\rightarrow t_{\mathrm{out%
}}}\approx \frac{J_{(d)}}{(2\pi )^{d-1}}\int_{\mathbf{p\in }D\left( t\right)
}d\mathbf{p}N_{n}^{\mathrm{cr}}\left\vert \left\langle P_{x}\left( t_{%
\mathrm{out}}\right) \right\rangle \right\vert ,  \notag \\
&&\left. \langle T_{ll}(t)\rangle \right\vert _{t>t_{\mathrm{out}}}\approx
\left. \langle T_{ll}(t)\rangle ^{p}\right\vert _{t\rightarrow t_{\mathrm{out%
}}}\approx \frac{J_{(d)}}{(2\pi )^{d-1}}\int_{\mathbf{p\in }D\left( t\right)
}d\mathbf{p}N_{n}^{\mathrm{cr}}\left\vert \left\langle P_{x}\left( t_{%
\mathrm{out}}\right) \right\rangle \right\vert ^{-1}p_{l}^{2},\ \;l=2,...,D.
\label{4.36b}
\end{eqnarray}

Using the universal form of the differential number of created pairs,
{\large \ }$N_{n}^{\mathrm{cr}}\approx N_{n}^{\mathrm{uni}}${\large , }given
by Eq.~(\ref{uni2}), making variable change (\ref{uni2b}), and performing
the integration over $p_{\bot },$\ we finally obtain from Eq.~(\ref{4.36b})
the new result. At the final time instant EMT components have the following
universal behavior:{\large \ }%
\begin{eqnarray}
&&\left. \langle T_{00}(t)\rangle \right\vert _{t>t_{\mathrm{out}}}\approx
\left. \langle T_{11}(t)\rangle \right\vert _{t>t_{\mathrm{out}}}\approx
\left. \langle T_{11}(t)\rangle ^{p}\right\vert _{t\rightarrow t_{\mathrm{out%
}}}  \notag \\
&&\ \approx \frac{J_{(d)}}{(2\pi )^{d-1}}\int_{t_{\mathrm{in}}}^{t_{\mathrm{%
out}}}dt\left[ U\left( t_{\mathrm{out}}\right) -U\left( t\right) \right] %
\left[ eE\left( t\right) \right] ^{d/2}\exp \left[ -\pi \frac{m^{2}}{%
eE\left( t\right) }\right] ,  \notag \\
&&\left. \langle T_{ll}(t)\rangle \right\vert _{t>t_{\mathrm{out}}}\approx
\left. \langle T_{ll}(t)\rangle ^{p}\right\vert _{t\rightarrow t_{\mathrm{out%
}}}\approx \frac{J_{(d)}}{(2\pi )^{d}}\int_{t_{\mathrm{in}}}^{t_{\mathrm{out}%
}}\frac{dt\left[ eE\left( t\right) \right] ^{d/2+1}}{\left[ U\left( t_{%
\mathrm{out}}\right) -U\left( t\right) \right] }\exp \left[ -\pi \frac{m^{2}%
}{eE\left( t\right) }\right] .  \label{4.36c}
\end{eqnarray}

The quantity $\left. \langle T_{00}(t)\rangle \right\vert _{t>t_{\mathrm{out}%
}}$ is the mean energy density of pairs created at any time instant $t$ with
zero longitudinal kinetic momentum and then accelerated to final
longitudinal kinetic momenta from zero to its maximum $\Delta U$. The
quantity $\left. \langle T_{11}(t)\rangle \right\vert _{t>t_{\mathrm{out}}}/2%
\tilde{n}^{\mathrm{cr}}$ is the mean kinetic momentum per particle at the
time instant $t_{\mathrm{out}}$. The energy density $\left. \langle
T_{00}(t)\rangle \right\vert _{t>t_{\mathrm{out}}}$ is equal to the pressure
$\left. \langle T_{11}(t)\rangle \right\vert _{t>t_{\mathrm{out}}}$ along
the direction of the electric field at the time instant $t_{\mathrm{out}}$.
This equality is a natural equation of state for noninteracting particles
accelerated by an electric field to relativistic velocities.

In particular, for fields admitting exactly solvable cases (these fields are
given by Eqs.~(\ref{2.8}), (\ref{t7}), and (\ref{ns4.0})), we find from Eq.~(%
\ref{4.36c}) [also recall the various definitions in Eq.~(\ref{4.19a})] the
following.

(i) For\ Sauter-like\ field:%
\begin{eqnarray}
&&\left. \langle T_{00}(t)\rangle ^{p}\right\vert _{t\rightarrow t_{\mathrm{%
out}}}\approx \left. \langle T_{11}(t)\rangle ^{p}\right\vert _{t\rightarrow
t_{\mathrm{out}}}\approx eE_{0}r^{\mathrm{cr}}T_{\mathrm{S}}^{2}\left[
\delta -G\left( \frac{d}{2},\frac{\pi m^{2}}{eE_{0}}\right) \right] ,  \notag
\\
&&\left. \langle T_{ll}(t)\rangle ^{p}\right\vert _{t\rightarrow t_{\mathrm{%
out}}}\approx \frac{r^{\mathrm{cr}}}{2\pi }\left[ \sqrt{\pi }\Psi \left(
\frac{1}{2},2-\frac{d}{2};\frac{\pi m^{2}}{eE_{0}}\right) +G\left( \frac{d}{2%
}-1,\frac{\pi m^{2}}{eE_{0}}\right) \right] ;  \label{4.38b}
\end{eqnarray}

(ii) For $T$-constant field:
\begin{eqnarray}
&&\left. \langle T_{00}(t)\rangle ^{p}\right\vert _{t\rightarrow t_{\mathrm{%
out}}}\approx \left. \langle T_{11}(t)\rangle ^{p}\right\vert _{t\rightarrow
t_{\mathrm{out}}}\approx eE_{0}r^{\mathrm{cr}}\left( t_{\mathrm{out}}-t_{%
\mathrm{in}}\right) ^{2},  \notag \\
&&\left. \langle T_{ll}(t)\rangle ^{p}\right\vert _{t\rightarrow t_{\mathrm{%
out}}}\approx \pi ^{-1}r^{\mathrm{cr}}\ln \left[ \sqrt{eE_{0}}\left( t_{%
\mathrm{out}}-t_{\mathrm{in}}\right) \right] ;\;  \label{4.37}
\end{eqnarray}

(iii) For the peak field:%
\begin{eqnarray}
&&\left. \langle T_{00}(t)\rangle ^{p}\right\vert _{t\rightarrow t_{\mathrm{%
out}}}\approx \left. \langle T_{11}(t)\rangle ^{p}\right\vert _{t\rightarrow
t_{\mathrm{out}}}\approx eE_{0}r^{\mathrm{cr}}\left[ k_{2}^{-1}+k_{1}^{-1}%
\right]  \notag \\
&&\times \left\{ \left[ k_{2}^{-1}-k_{1}^{-1}\right] G\left( \frac{d}{2}+1,%
\frac{\pi m^{2}}{eE_{0}}\right) +k_{1}^{-1}G\left( \frac{d}{2},\frac{\pi
m^{2}}{eE_{0}}\right) \right\} ,  \notag \\
&&\left. \langle T_{ll}(t)\rangle ^{p}\right\vert _{t\rightarrow t_{\mathrm{%
out}}}\approx \frac{r^{\mathrm{cr}}}{2\pi }\left[ G\left( \frac{d}{2}-1,%
\frac{\pi m^{2}}{eE_{0}}\right) +\frac{k_{2}}{k_{1}}G\left( \frac{d}{2},%
\frac{\pi m^{2}}{eE_{0}}\right) \right] ,\;  \label{4.38}
\end{eqnarray}%
where $l=2,...,D$. Densities (\ref{4.38}) correspond to the case of an
exponentially decaying field as $k_{1}^{-1}\rightarrow 0$.

Note that using the differential mean numbers of created pairs given by
Eqs.~(\ref{asy2}), (\ref{asy4}), and (\ref{4.10}) for the exactly solvable
cases, we obtain from Eq.~(\ref{4.36b})~literally expressions{\large \ }(\ref%
{4.37}) (earlier obtained in Refs.~\cite{GG06-08,GavGitY12}), (\ref{4.38b}),
and (\ref{4.38}). It is an independent confirmation of universal form (\ref%
{4.36c}).{\large \ }We stress that Eqs.~(\ref{4.38b}) and (\ref{4.38}) are
first obtained in this article.

It should be noted that the densities $\left. \langle j^{1}(t)\rangle
^{p}\right\vert _{t\rightarrow t_{\mathrm{out}}}$ and $\left. \langle T_{\mu
\mu }(t)\rangle ^{p}\right\vert _{t\rightarrow t_{\mathrm{out}}}$ are formed
over the entire time interval $t_{\mathrm{out}}-t_{\mathrm{in}}$ of the
field duration. All these densities are growing functions of the increment
of the longitudinal kinetic momentum. However, they differ, in particular,
because switching on and off conditions of the corresponding electric fields
are different.

In what follows we show that some universal behavior of the densities{\large %
\ }$\langle j^{1}(t)\rangle ^{p}${\large \ }and{\large \ }$\langle T_{\mu
\mu }(t)\rangle ^{p}${\large \ }can be derived from general forms~(\ref{4.33}%
) for any large difference{\large \ }$t-t_{\mathrm{in}}${\large ,} even if%
{\large \ }$t-t_{\mathrm{in}}\ll t_{\mathrm{out}}-t_{\mathrm{in}}${\large . }%
We begin the demonstration of this fact with the case of a finite interval
of time when the electric field potential can be approximated by a potential
of a constant electric field (\ref{4.39}). At the same time, we assume that $%
\left\langle P_{x}\left( t\right) \right\rangle $ satisfies condition (\ref%
{4.31a}) at the time $t$. It is convenient to compare the cases of $T$%
-constant and exponentially decaying fields, which both are abruptly
switching on but their ways of switching off may be different.

In the case of an exponentially decaying field, the functions $^{\pm
}\varphi _{n}\left( t\right) $ in Eq.~(\ref{4.33}) are given by the second
line in Eq.~(\ref{i.4.1}) and approximation (\ref{4.39}) holds if $k_{2}t\ll
1$. Then $\left\vert \left\langle P_{x}\left( t\right) \right\rangle
\right\vert \ll \left\vert \pi _{2}\right\vert $. To obtain functions $%
\;^{\pm }\varphi _{n}\left( t\right) $ in such an approximation we use first
the asymptotic representation for the confluent hypergeometric function $%
\Phi \left( a,c;\eta \right) $ via the Weber parabolic cylinder functions
(WPCFs) for large $\eta $ and $c$ with fixed $a$ and $\tau =\eta /c\sim 1$,
given by Eq.~(13.8.4) in \cite{DLMF}. Assuming then $\left\vert \tau
-1\right\vert \sim 1$ and using asymptotic expansions of WPCFs one finds $%
\Phi \left( a,c;\eta \right) \approx \left( 1-\tau \right) ^{-a}$ for $%
1-\tau >0$. Thus, we obtain%
\begin{equation}
\rho \left( t\right) =\left[ V_{\left( d-1\right) }\left\vert \left\langle
P_{x}\left( t\right) \right\rangle \right\vert \right] ^{-1}.  \label{4.40}
\end{equation}%
In the range {\large \ }$D\left( t\right) $, the distribution $N_{n}^{%
\mathrm{cr}}$ is approximately given by Eq.~(\ref{asy4}). Finally we obtain%
\begin{eqnarray}
&\langle j^{1}(t)\rangle ^{p}\approx &2er^{\mathrm{cr}}\Delta t,  \notag \\
&\langle T_{00}(t)\rangle ^{p}\approx &\left. \langle T_{11}(t)\rangle
^{p}\right\vert \approx eE_{0}r^{\mathrm{cr}}\Delta t^{2},  \notag \\
&\langle T_{ll}(t)\rangle ^{p}\approx &\pi ^{-1}r^{\mathrm{cr}}\ln \left(
\sqrt{eE_{0}}\Delta t\right) \;\mathrm{if}\;l=2,...,D,  \label{4.41}
\end{eqnarray}%
where $\Delta t=t-t_{\mathrm{in}}$ is the duration time of a constant field.
In this case $t_{\mathrm{in}}=0$.

The field potential of the $T$-constant field (\ref{t7}) has the form (\ref%
{4.39}) in the intermediate region{\Huge \ }$\mathrm{II}$. For{\Large \ }%
sufficiently large times $t<t_{\mathrm{out}}$,\ when the longitudinal
kinetic momentum belongs to the range {\large \ }$D\left( t\right) $, the
distribution $N_{n}^{\mathrm{cr}}$ is approximately given by Eq.~(\ref{asy4}%
). In this case, exact expressions for the functions $\;^{+}\varphi
_{n}\left( t\right) $, see Eq.~(26)~in Ref.~\cite{GavG96a}, and similar
expressions for the functions $\;^{-}\varphi _{n}\left( t\right) $ can be
approximated as the following WPCFs:%
\begin{eqnarray}
&&\ ^{+}\varphi _{n}\left( t\right) \approx V_{\left( d-1\right)
}^{-1/2}CD_{-1-\rho }[(1+i)\xi ],\ \;^{-}\varphi _{n}\left( t\right) \approx
V_{\left( d-1\right) }^{-1/2}CD_{\rho }[(1-i)\xi ],  \notag \\
&&\xi =\left( eE_{0}t-p_{x}\right) \left( eE_{0}\right) ^{-1/2},\ \ \
C=\left( 2eE_{0}\right) ^{-1/2}\exp \left( -\pi \lambda _{0}/8\right) \,.
\label{4.42}
\end{eqnarray}%
Then we find from Eq.~(\ref{4.33}) that the densities $\langle
j^{1}(t)\rangle ^{p}$ and $\langle T_{\mu \mu }(t)\rangle ^{p}$ have the
same form (\ref{4.41}) with $t_{\mathrm{in}}=-T/2$.

Note that the above results are obtained by using functions $^{\pm }\varphi
_{n}\left( t\right) $, which have in and out asymptotics at $t_{\mathrm{out}%
} $.{\large \ }Nevertheless, these results show also that densities (\ref%
{4.41}) are not affected by evolution of the functions $^{\pm }\varphi
_{n}\left( t\right) $\ from $t$ to $t_{\mathrm{out}}$ in the range $p\in
D\left( t\right) $, assuming that the corresponding electric field exists
during a macroscopically large time period $\Delta t$, satisfying Eq.~(\ref%
{svf1}). This fact is closely related with a characteristic property of the
kernel of integrals (\ref{4.33}){\large , }which will be derived from a
universal form of the total density of created pairs given by Eq.~(\ref{uni2}%
).{\large \ }Let $t_{\mathrm{out}}^{\prime }<t_{\mathrm{out}}$\ be another
possible final time instant. Then%
\begin{equation}
\tilde{n}^{\mathrm{cr}}\left( t_{\mathrm{out}}^{\prime }\right) \approx
\frac{J_{(d)}}{(2\pi )^{d-1}}\int_{t_{\mathrm{in}}}^{t_{\mathrm{out}%
}^{\prime }}dt\left[ eE\left( t\right) \right] ^{d/2}\exp \left\{ -\pi \frac{%
m^{2}}{eE\left( t\right) }\right\}  \label{uni8}
\end{equation}%
Equation~(\ref{uni8}) corresponds to the assumption that in the range $p\in
\ D\left( t_{\mathrm{out}}^{\prime }\right) \subset D\left( t_{\mathrm{out}%
}\right) $ the electric field is switched on at $t_{\mathrm{in}}$\ and
switched off at $t_{\mathrm{out}}^{\prime }$. Then instead of functions\ \ $%
^{\zeta }\psi _{n}\left( x\right) $\ satisfying the eigenvalue problem (\ref%
{t4.ba}), we have to use solutions of the following eigenvalue problem%
\begin{equation*}
H\left( t\right) \ ^{\zeta }\psi _{n}^{\left( t_{\mathrm{out}}^{\prime
}\right) }\left( x\right) =\ ^{\zeta }\varepsilon _{n}\ ^{\zeta }\psi
_{n}^{\left( t_{\mathrm{out}}^{\prime }\right) }\left( x\right) \,,\ \ t\in %
\left[ t_{\mathrm{out}}^{\prime },+\infty \right) \,,\ ^{\zeta }\varepsilon
_{n}=\zeta p_{0}\left( t_{\mathrm{out}}^{\prime }\right) \,.
\end{equation*}%
Using the representation%
\begin{equation*}
\ ^{\zeta }\psi _{n}^{\left( t_{\mathrm{out}}^{\prime }\right) }\left(
x\right) =\left[ i\partial _{t}+H\left( t\right) \right] \gamma ^{0}\exp
\left( i\mathbf{pr}\right) \ ^{\zeta }\varphi _{n}^{\left( t_{\mathrm{out}%
}^{\prime }\right) }\left( t\right) v_{\chi ,\sigma }
\end{equation*}%
we obtain%
\begin{eqnarray}
&&\ ^{\zeta }\varphi _{n}^{\left( t_{\mathrm{out}}^{\prime }\right) }\left(
t\right) =\ ^{\zeta }N^{\left( t_{\mathrm{out}}^{\prime }\right) }\exp \left[
-i\zeta p_{0}\left( t_{\mathrm{out}}^{\prime }\right) \left( t-t_{\mathrm{out%
}}^{\prime }\right) \right] \,,\ \ t\in \left[ t_{\mathrm{out}}^{\prime
},+\infty \right) \,,  \notag \\
&&\ ^{\zeta }N^{\left( t_{\mathrm{out}}^{\prime }\right) }=\left(
2p_{0}\left( t_{\mathrm{out}}^{\prime }\right) \left\{ p_{0}\left( t_{%
\mathrm{out}}^{\prime }\right) -\chi \zeta \left[ p_{x}-U\left( t_{\mathrm{%
out}}^{\prime }\right) \right] \right\} V_{\left( d-1\right) }\right)
^{-1/2}\ .  \label{uni9}
\end{eqnarray}%
Thus, the leading contribution to the function $S^{p}(x,x^{\prime })$
[defined by Eq.~(\ref{3.25})] at $t^{\prime }\sim $ $t<t_{\mathrm{out}%
}^{\prime }$ can be expressed via $\ ^{\zeta }\psi _{n}^{\left( t_{\mathrm{%
out}}^{\prime }\right) }\left( x\right) $ as follows%
\begin{equation}
S^{p}(x,x^{\prime })\approx -i\sum_{\sigma ,\mathbf{p\in }D\left( t\right)
}N_{n}^{\mathrm{cr}}\left[ \ ^{+}{\psi }_{n}^{\left( t_{\mathrm{out}%
}^{\prime }\right) }(x)\ ^{+}{\bar{\psi}}_{n}^{\left( t_{\mathrm{out}%
}^{\prime }\right) }(x^{\prime })-\,\ ^{-}{\psi }_{n}^{\left( t_{\mathrm{out}%
}^{\prime }\right) }(x)\ ^{-}{\bar{\psi}}_{n}^{\left( t_{\mathrm{out}%
}^{\prime }\right) }(x^{\prime })\right] \,.  \label{uni10}
\end{equation}%
Then $\rho \left( t\right) $\ in Eq.~(\ref{4.33}) can be represented as%
\begin{equation*}
\rho \left( t\right) =2\left\vert \left\langle P_{x}\left( t\right)
\right\rangle \right\vert \left\{ \left. \left\vert \ ^{+}\varphi
_{n}^{\left( t_{\mathrm{out}}^{\prime }\right) }\left( t\right) \right\vert
^{2}\right\vert _{\chi =+1}+\left. \left\vert \ ^{-}\varphi _{n}^{\left( t_{%
\mathrm{out}}^{\prime }\right) }\left( t\right) \right\vert ^{2}\right\vert
_{\chi =-1}\right\} .
\end{equation*}%
Taking into account Eq.~(\ref{uni9}), we can see that Eq.~(\ref{4.40}) holds
for any large time difference $t-t_{\mathrm{in}}$. Using the universal form
of the differential numbers of created pairs, \ $N_{n}^{\mathrm{cr}}\approx
N_{n}^{\mathrm{uni}}$, given by Eq.~(\ref{uni2}), changing the variable
according to Eq.~(\ref{uni2b}), and performing the integration over $p_{\bot
}$, we find from Eq.~(\ref{4.33}) that the vacuum mean values of current and
EMT components have the following universal behavior for any large
difference $t-t_{\mathrm{in}}$:%
\begin{eqnarray}
&\langle j^{1}(t)\rangle ^{p}\approx &2e\tilde{n}^{\mathrm{cr}}\left(
t\right) ,  \notag \\
&\langle T_{00}(t)\rangle ^{p}\approx &\langle T_{11}(t)\rangle ^{p}\approx
\frac{J_{(d)}}{(2\pi )^{d-1}}\int_{t_{\mathrm{in}}}^{t}dt^{\prime }\left[
U\left( t\right) -U\left( t^{\prime }\right) \right] \left[ eE\left(
t^{\prime }\right) \right] ^{d/2}\exp \left[ -\frac{\pi m^{2}}{eE\left(
t^{\prime }\right) }\right] ,  \notag \\
&\langle T_{ll}(t)\rangle ^{p}\approx &\frac{J_{(d)}}{(2\pi )^{d}}\int_{t_{%
\mathrm{in}}}^{t}\frac{dt^{\prime }\left[ eE\left( t^{\prime }\right) \right]
^{d/2+1}}{\left[ U\left( t\right) -U\left( t^{\prime }\right) \right] }\exp %
\left[ -\frac{\pi m^{2}}{eE\left( t^{\prime }\right) }\right] ,\;l=2,...,D\ .
\label{uni11}
\end{eqnarray}%
{\large \ }Here $\tilde{n}^{\mathrm{cr}}\left( t\right) $\ is given by Eq.~(%
\ref{uni8}). In particular, when $t=t_{\mathrm{out}}$\ one obtains Eq.~(\ref%
{4.36a}) and (\ref{4.36c}).

The obtained results show that the scale $\Delta t_{\mathrm{st}}^{\mathrm{m}%
} $ plays the role of the stabilization time for the densities $\langle
j^{1}(t)\rangle ^{p}$ and $\langle T_{\mu \mu }(t)\rangle ^{p}$. The
characteristic parameter $m^{2}/eE_{0}$ can be represented as the ratio of
two characteristic lengths: $c^{3}m^{2}/\hslash eE_{0}=\left( c\Delta t_{%
\mathrm{st}}/\Lambda _{C}\right) ^{2}$, where $\Lambda _{C}=\hbar /mc$ is
the Compton wavelength. In strong electric fields, $\left( c\Delta t_{%
\mathrm{st}}/\Lambda _{C}\right) ^{2}\lesssim 1$, inequality (\ref{svf2}) is
simplified to the form $\Delta t/\Delta t_{\mathrm{st}}\gg 1$, in which the
Compton wavelength is absent. We see that the scale $\Delta t_{\mathrm{st}}$
plays the role of the stabilization time for a strong electric field. This
means that $\Delta t_{\mathrm{st}}$\ is a characteristic time scale which
allows us to distinguish fields that have microscopic or macroscopic time
change. It plays a role similar to that of the Compton wavelength in the
case of a weak field. Therefore, calculations in a $T$-constant field are
quite representative for a large class of slowly varying electric fields.

In what follows we use the example of the $T$-constant field to consider the%
{\large \ }contributions{\large \ }$\mathrm{Re}\langle j^{\mu }(t)\rangle
^{c}${\large \ }and{\large \ }$\mathrm{Re}\langle T_{\mu \nu }(t)\rangle
^{c} ${\large \ }to the mean values of the current density $\langle j^{\mu
}(t)\rangle $\ and the EMT $\langle T_{\mu \nu }(t)\rangle $, given by Eqs.~(%
\ref{A1.4}). Note that the mean current density $\,\langle j^{\mu
}(t)\rangle $ and the physical part of the mean value $\langle T_{\mu \nu
}(t)\rangle $ are zero for any $t<t_{\mathrm{in}}$. For $t>t_{\mathrm{in}}$,
we are interested in these mean values only for a large time periods $\Delta
t=t-t_{\mathrm{in}}$ satisfying Eq.~ (\ref{svf1}). In this case, the
longitudinal kinetic momentum belongs to the range (\ref{4.31a}) and
distributions $N_{n}^{\mathrm{cr}}$ are approximated by Eq.~(\ref{asy4}).
Using approximation (\ref{4.42}), the functions $\;_{-}\varphi _{n}\left(
t\right) $, given by Eq.~(25)~in Ref.~\cite{GavG96a},{\large \ }and similar
functions $\;_{+}\varphi _{n}\left( t\right) $ can be taken in the following
form%
\begin{equation}
\ _{-}\varphi _{n}\left( t\right) =V_{\left( d-1\right) }^{-1/2}CD_{-1-\rho
}[-(1+i)\xi ],\;_{+}\varphi _{n}\left( t\right) =V_{\left( d-1\right)
}^{-1/2}CD_{\rho }[-(1-i)\xi ].  \label{4.44}
\end{equation}%
In the same approximation, the causal propagator $S^{c}(x,x^{\prime })$ (\ref%
{3.22}) can be calculated using solutions $\ ^{\pm }\psi _{n}(x)$ and $\
_{\pm }\psi _{n}(x)$ with scalar functions given by Eqs.~(\ref{4.42}) and (%
\ref{4.44}) in the range (\ref{4.31a}). It can be shown that the main
contributions to $\mathrm{Re}\,\langle j^{\mu }(t)\rangle ^{c}$, $\langle
j^{2}(t)\rangle $ and $\mathrm{Re}\,\langle T_{\mu \mu }(t)\rangle ^{c}$ are
formed in the range (\ref{4.31a}) for a large time period $\Delta t$. It is
important that these contributions are independent of the interval $\Delta t$%
, that is, the densities $\mathrm{Re}\,\langle j^{\mu }(t)\rangle ^{c}$, $%
\langle j^{2}(t)\rangle $, and $\mathrm{Re}\,\langle T_{\mu \mu }(t)\rangle
^{c}$ are local quantities describing only vacuum polarization\ effects.
Then we integrate in Eq.~(\ref{3.22}) over all the momenta. Thus, we see
that in the case under consideration, the propagator $S^{c}(x,x^{\prime })$
can be approximated by the propagator in a constant uniform electric field.

The propagator $S^{c}(x,x^{\prime })$ in a constant uniform electric field
can be represented as the Fock-Schwinger proper-time integral:%
\begin{equation}
S^{c}(x,x^{\prime })=(\gamma P+m)\Delta ^{c}(x,x^{\prime }),\;\;\Delta
^{c}(x,x^{\prime })=\int_{0}^{\infty }f(x,x^{\prime },s)ds\,,  \label{4.45}
\end{equation}%
see \cite{Nikis70a} and \cite{GGG98}, where the Fock-Schwinger kernel $%
f(x,x^{\prime },s)$ reads%
\begin{align}
& f(x,x^{\prime },s)=\exp \left( i\frac{e}{2}\sigma ^{\mu \nu }F_{\mu \nu
}s\right) f^{(0)}(x,x^{\prime },s)\,,\ \ f^{(0)}(x,x^{\prime },s)=-\frac{%
eE_{0}s^{-d/2+1}}{(4\pi i)^{d/2}\sinh (eE_{0}s)}  \notag \\
& \times \exp \left[ -i\left( e\Lambda +m^{2}s\right) +\frac{1}{4i}%
(x-x^{\prime })eF\coth (eFs)(x-x^{\prime })\right] .  \notag
\end{align}%
Here $\coth (eFs)$ is the matrix with the components $[\coth (eFs)]^{\mu
}{}_{\nu }$, $F_{\mu \nu }=E_{0}\left( \delta _{\mu }^{0}\delta _{\nu
}^{1}-\delta _{\mu }^{1}\delta _{\nu }^{0}\right) $, and $\Lambda
=(t+t^{\prime })(x_{1}-x_{1}^{\prime })E_{0}/2$; see \cite{Schwinger51,F37}%
.\

It is easy to see that $\langle j^{1}\left( t\right) \rangle ^{c}=0$, as
should be expected due to the translational symmetry. If $d=3$ there is a
transverse vacuum-polarization current,
\begin{equation}
\langle j^{2}(t)\rangle =\pm \frac{e^{2}}{4\pi ^{3/2}}\gamma \left( \frac{1}{%
2},\frac{\pi m^{2}}{eE_{0}}\right) E_{0},  \label{emt2d}
\end{equation}%
for each $\pm $ fermion, \cite{GavGitY12} (see, as well, Ref.~\cite{DoraM11}%
), where $\gamma \left( 1/2,x\right) $ is the incomplete gamma function.
Note that the transverse current of created particles is absent, $\langle
j^{2}(t)\rangle =0$ if $t>t_{\mathrm{out}}$. The factor in the front of $%
E_{0}$ in Eq. (\ref{emt2d}) can be considered as a nonequilibrium Hall
conductivity for large duration of the electric field. In the presence of
both $\pm $ fermions in a model, $\langle j^{2}(t)\rangle =0$ for any $t$.

Using Eq.~(\ref{4.45}), we obtain components of the EMT for the $T$-constant
field in the following form
\begin{align}
& \mathrm{Re}\langle T_{00}(t)\rangle ^{c}=-\mathrm{Re}\langle
T_{11}(t)\rangle ^{c}=E_{0}\frac{\partial \mathrm{Re}\mathcal{L}\left[ E_{0}%
\right] }{\partial E_{0}}-\mathrm{Re}\mathcal{L}\left[ E_{0}\right] \,,
\notag \\
& \mathrm{Re}\langle T_{ll}(t)\rangle ^{c}=\mathrm{Re}\mathcal{L}\left[ E_{0}%
\right] \,,\;l=2,...,D,  \label{4.46a}
\end{align}%
where
\begin{equation}
\mathcal{L}\left[ E_{0}\right] =\frac{1}{2}\int_{0}^{\infty }\frac{ds}{s}%
\mathrm{tr}f(x,x,s),\ \mathrm{tr}f(x,x,s)=2^{[d/2]}\cosh
(eE_{0}s)f^{(0)}(x,x,s).  \label{ELaa}
\end{equation}

The quantity $\mathcal{L}\left[ E_{0}\right] $ can be identified with a
nonrenormalized one-loop effective Euler-Heisenberg Lagrangian of the Dirac
field in a uniform constant electric field $E_{0}$. Note that components $%
\mathrm{Re}\langle T_{\mu \nu }(t)\rangle ^{c}$ do not depend on the time
duration $\Delta t$\ of the $T$-constant field if $\Delta t$ is sufficiently
large.

This result can be generalized to the case of an arbitrarily slowly varying
electric field{\large . }To this end we divide as before the finite interval
$\left( t_{\mathrm{in}}^{eff},t_{\mathrm{out}}^{eff}\right] $\ into $M$\
intervals{\large \ }$\Delta t_{i}=t_{i+1}-t_{i}>0${\large , }such that Eq.~(%
\ref{svf1}) holds true for each of them. That allows us to treat the
electric field as approximately constant within each{\large \ }interval,%
{\large \ }$\overline{E(t)}\approx \overline{E}(t_{i})${\large \ for }$t\in
\left( t_{i},t_{i+1}\right] ${\large . }In each such interval, we obtain
expressions similar to the ones{\large \ }(\ref{4.46a}) and (\ref{ELaa}),
where the electric field $E_{0}$\ has to be substituted by{\large \ }$%
\overline{E}(t_{i})$. Then{\large \ }components of the EMT for an
arbitrarily slowly varying strong electric field $E\left( t\right) $\ in the
leading-term approximation can be represented as
\begin{align}
& \mathrm{Re}\langle T_{00}(t)\rangle ^{c}=-\mathrm{Re}\langle
T_{11}(t)\rangle ^{c}=E\left( t\right) \frac{\partial \mathrm{Re}\mathcal{L}%
\left[ E\left( t\right) \right] }{\partial E\left( t\right) }-\mathrm{Re}%
\mathcal{L}\left[ E\left( t\right) \right] \,,  \notag \\
& \mathrm{Re}\langle T_{ll}(t)\rangle ^{c}=\mathrm{Re}\mathcal{L}\left[
E\left( t\right) \right] \,,\;l=2,...,D,  \label{4.46b}
\end{align}%
where
\begin{eqnarray}
&&\mathcal{L}\left[ E\left( t\right) \right] =\frac{1}{2}\int_{0}^{\infty }%
\frac{ds}{s}\mathrm{tr}\tilde{f}(x,x,s),\ \mathrm{tr}\tilde{f}%
(x,x,s)=2^{[d/2]}\cosh \left[ eE\left( t\right) s\right] \tilde{f}%
^{(0)}(x,x,s),  \notag \\
&&\tilde{f}^{(0)}(x,x,s)=-\frac{eE\left( t\right) s^{-d/2+1}\exp \left(
-im^{2}s\right) }{(4\pi i)^{d/2}\sinh \left[ eE\left( t\right) s\right] }.
\label{4.46c}
\end{eqnarray}%
Note that{\large \ }$\mathcal{L}\left[ E\left( t\right) \right] ${\large \ }%
evolves in time due to the time dependence of the field $E\left( t\right) $%
{\large .}

The quantity $\mathcal{L}\left[ E\left( t\right) \right] $ describes the
vacuum polarization. The quantities (\ref{4.46b}) are divergent due to the
real part of the effective Lagrangian (\ref{4.46c}), which is ill defined.
This real part must be regularized and renormalized. In low dimensions, $%
d\leq 4$, $\mathrm{Re}\mathcal{L}\left[ E\left( t\right) \right] $ can be
regularized in the proper-time representation and renormalized by the
Schwinger renormalizations of the charge and the electromagnetic field \cite%
{Schwinger51}. In particular, for $d=4$, the renormalized effective
Lagrangian $\mathcal{L}_{ren}\left[ E\left( t\right) \right] $ is
\begin{equation}
\mathcal{L}_{ren}\left[ E\left( t\right) \right] =\int_{0}^{\infty }\frac{%
ds\exp \left( -im^{2}s\right) }{8\pi ^{2}s}\left\{ \frac{eE\left( t\right)
\coth \left[ eE\left( t\right) s\right] }{s}-\frac{1}{s^{2}}-\frac{\left[
eE\left( t\right) s\right] ^{2}}{3}\right\} .  \label{HEL1}
\end{equation}%
In higher dimensions, $d>4$, a different approach is required. One can give
a precise meaning and calculate the one-loop effective action using
zeta-function regularization, see details in Ref.~\cite{GavGitY12}. If we
are interested in the case of a very strong field, $m^{2}/eE\left( t\right)
\ll 1$, then
\begin{equation}
\mathrm{Re}\mathcal{L}_{ren}\left[ E\left( t\right) \right] \sim \left\{
\begin{array}{ll}
\left[ eE\left( t\right) \right] ^{d/2}\,, & d\neq 4k\, \\
\left[ eE\left( t\right) \right] ^{d/2}\ln \left[ eE\left( t\right) /M^{2}%
\right] \,, & d=4k\,,\ k\in \mathbb{N}\,%
\end{array}%
\right. ,  \label{4.47}
\end{equation}%
where the quantity $M$ is a renormalization scale. In the framework of the
on-shell renormalization of a massive theory, we have to set $M=m$. Making
the same renormalization for $\langle T_{\mu \mu }(t)\rangle ^{c}$, we can
see that for the renormalized EMT the following relations hold true
\begin{align}
& \mathrm{Re}\,\langle T_{00}(t)\rangle _{ren}^{c}=-\mathrm{Re}\langle
T_{11}(t)\rangle _{ren}^{c}=E\left( t\right) \frac{\partial \mathrm{Re}%
\mathcal{L}_{ren}\left[ E\left( t\right) \right] }{\partial E\left( t\right)
}-\mathrm{Re}\mathcal{L}_{ren}\left[ E\left( t\right) \right] ,  \notag \\
& \mathrm{Re}\,\langle T_{ll}(t)\rangle _{ren}^{c}=\mathrm{Re}\mathcal{L}%
_{ren}\left[ E\left( t\right) \right] \,,\quad \ l=2,3,\dots ,D.
\label{L-ren}
\end{align}%
In the strong-field case, the leading contributions to the renormalized EMT
are
\begin{equation}
\mathrm{Re}\,\langle T_{\mu \mu }(t)\rangle _{ren}^{c}\sim \left\{
\begin{array}{l}
\left[ eE\left( t\right) \right] ^{d/2}\,,\quad d\neq 4k\, \\
\left[ eE\left( t\right) \right] ^{d/2}\ln \left[ eE\left( t\right) /M^{2}%
\right] ,\quad d=4k%
\end{array}%
\right. .  \label{emt-lead}
\end{equation}

The final form of the vacuum mean components of the EMT are
\begin{equation}
\langle T_{\mu \mu }(t)\rangle _{ren}=\mathrm{Re}\,\langle T_{\mu \mu
}(t)\rangle _{ren}^{c}+\mathrm{Re}\,\langle T_{\mu \mu }(t)\rangle ^{p},
\label{emt-ren}
\end{equation}%
where the {components }$\mathrm{Re}\,\langle T_{\mu \mu }(t)\rangle
_{ren}^{c}${\ and $\mathrm{Re}\,\langle T_{\mu \mu }(t)\rangle ^{p}$ are
given by }Eqs.~(\ref{L-ren}) and (\ref{uni11}), respectively.{\ }For $%
t<t_{in}$ and $t>t_{out}$ the electric field is absent such that $\mathrm{Re}%
\langle T_{\mu \mu }(t)\rangle _{ren}^{c}=0$.

On the right-hand side of Eq.~(\ref{emt-ren}), the term $\mathrm{Re}%
\,\langle T_{\mu \mu }(t)\rangle ^{p}$ represents contributions due to the
vacuum instability, whereas the term $\mathrm{Re}\,\langle T_{\mu \mu
}(t)\rangle _{ren}^{c}$ represents\textrm{\ }vacuum polarization effects.
For weak electric fields, $m^{2}/eE_{0}\gg 1,$ contributions due to the
vacuum instability are exponentially small, so that the vacuum polarization
effects play the principal role. For strong electric fields, $%
m^{2}/eE_{0}\ll 1$, the energy density of the vacuum polarization $%
Re\,\langle T_{00}(t)\rangle _{ren}^{c}$\ is negligible compared to the
energy density due to the vacuum instability $\langle T_{00}\left( t\right)
\rangle ^{p}$,%
\begin{equation}
\langle T_{\mu \mu }(t)\rangle _{ren}\approx \mathrm{Re}\,\langle T_{\mu \mu
}(t)\rangle ^{p}.  \label{emt-cr}
\end{equation}%
The latter density is formed on the whole time interval $t-t_{\mathrm{in}}$;%
{\large \ }however, dominant contributions are due to time intervals $\Delta
t_{i}$ with $m^{2}/e\overline{E}(t_{i})<1$\ and the large dimensionless
parameters $\sqrt{e\overline{E}(t_{i})}\Delta t_{i}$.

We note that the effective Lagrangian (\ref{4.46c}) and its renormalized
form $\mathcal{L}_{ren}\left[ E\left( t\right) \right] $\ coincide with
leading term approximation of derivative expansion results from
field-theoretic calculations obtained in Refs.~\cite{DunnH98,GusSh99} for%
{\large \ }$d=3${\large \ }and{\large \ }$d=4${\large . }In this
approximation the $S^{\left( 0\right) }$\ term of the Schwinger's effective
action, given by the expansion (\ref{uni7b}), has the form
\begin{equation}
S^{\left( 0\right) }[F_{\mu \nu }]=\int dx\mathcal{L}_{ren}\left[ E\left(
t\right) \right] .  \label{derexp}
\end{equation}

It should be stressed that unlike the authors of Refs.~\cite{DunnH98,GusSh99}%
, we derive Eq.~(\ref{4.46c}) and its renormalized form in the framework of
the general exact formulation of strong-field QED \cite{FGS,GavGT06}, using
the QED definition of the mean value of the EMT, given by Eq.~(\ref{A1.4}).%
{\large \ }Therefore we obtain $\mathcal{L}_{ren}\left[ E\left( t\right) %
\right] $ independently from the derivative expansion approach and the
obtained result holds true for any strong field under consideration.{\large %
\ }Moreover, it is proven that in this case not only the imaginary part of $%
S^{\left( 0\right) }$ but also its real part are given exactly by the
semiclassical WKB limit. It is clearly demonstrated that the imaginary part
of the effective action,{\large \ }$\mathrm{Im}S^{\left( 0\right) }$, is
related to the vacuum-to-vacuum transition probability $P_{v}$\ and can be
represented as an integral of $\mathcal{L}_{ren}\left[ E\left( t\right) %
\right] $ over the total field history, whereas the kernel of the real part
of this effective action,{\large \ }$\mathrm{Re}\mathcal{L}_{ren}\left[
E\left( t\right) \right] $, is related to the local EMT which defines the
vacuum polarization. Obtained results justify the derivative expansion as an
asymptotic expansion that can be useful to find the corrections for mean
values of the EMT components. We also note that some authors have argued
that the locally constant field approximation, which amounts to limiting
oneself to the leading contribution of the\ derivative expansion of the
effective action, allows for reliable results for electromagnetic fields of
arbitrary strength; cf., e.g., \cite{GalN83,GiesK17}.

Exemplarily focusing on the $T$-constant field, in the following we
demonstrate that under natural assumptions, the parameter $eE_{0}\Delta
t^{2} $ is limited. For $d>4$ an exact meaning of finite terms of the
effective Lagrangian (\ref{ELaa}) can be understood only from the
corresponding fundamental theory. Considering problems of high-energy
physics in $d=4,$ it is usually assumed that just from the beginning there
exists a uniform classical electric field with a given energy density. The
system of particles interacting with this field is closed; that is, the
total energy of the system is conserved. Under such an assumption, the pair
creation is a transient process and the applicability of the constant field
approximation is limited by the smallness of the backreaction, which implies
the following restriction from above:%
\begin{equation}
\left( \Delta t/\Delta t_{\mathrm{st}}\right) ^{2}\ll \frac{\pi ^{2}}{%
J\alpha }\,\exp \left( \pi \frac{c^{3}m^{2}}{\hslash eE_{0}}\right) ,
\label{4.48}
\end{equation}%
on time $\Delta t$ for a given electric field strength. Here $\alpha $ is
the fine structure constant and $J$ is the number of the spin degrees of
freedom, see \cite{GavG08}. Thus, there is a {range of} the parameters $%
E_{0} $ and $\Delta t$ where the approximation of the constant external
field is consistent. For QCD with a constant $SU(3)$ chromoelectric field $%
E_{0}^{a}$ ($a=1,\ldots ,8$) (during the period when the produced partons
can be treated as weakly coupled due to the property of asymptotic freedom
in QCD), and at low temperatures $\theta \ll q\sqrt{C_{1}}\Delta t,$ the
consistency restriction for the dimensionless parameter $q\sqrt{C_{1}}\Delta
t^{2}$ has the form $1\ll q\sqrt{C_{1}}\Delta t^{2}\ll \pi ^{2}/3q^{2}\,,$
where $q$ is the coupling constant and $C_{1}=E_{0}^{a}E_{0}^{a}$ is a
Casimir invariant for{\large \ }$SU(3)$.

The case of $d=3$ attracts attention in recent years. It is well known that
at certain conditions (the so-called charge neutrality point) electronic
excitations in the graphene monolayer behave as relativistic Dirac massless
fermions in $2+1$ dimensions, with the Fermi velocity $v_{F}\simeq 10^{6}$
\textrm{m/s} playing the role of the speed of light, see details in recent
reviews \cite{dassarma,VafVish14}. Then in the range of the applicability of
the Dirac model to the graphene physics, any electric field is strong. There
appears a time scale specific to graphene (and to similar nanostructures
with the Dirac fermions), $\Delta t_{\mathrm{st}}^{\mathrm{g}}=\left(
eE_{0}v_{F}/\hbar \right) ^{-1/2}\,$, which plays the role of the
stabilization time in the case under consideration. The generation of a mass
gap in the graphene band structure is an important fundamental and practical
problem under current research. In the presence of the mass gap $\Delta
\varepsilon =mv_{F}^{2}$, the stabilization condition has a general form%
\begin{equation}
\Delta t/\Delta t_{\mathrm{st}}^{\mathrm{gm}}\gg 1,\;\Delta t_{\mathrm{st}}^{%
\mathrm{gm}}=\Delta t_{\mathrm{st}}^{\mathrm{g}}\max \left\{ 1,(\Delta
\varepsilon )^{2}/\hslash v_{F}eE_{0}\right\} .  \label{4.51}
\end{equation}%
In this case, the strong-field condition reads $(\Delta \varepsilon
)^{2}/v_{F}\hbar eE_{0}\ll 1$. It has been shown \cite{lewkowicz10} that the
time scale $\Delta t_{\mathrm{st}}^{\mathrm{g}}$ appears for the
tight-binding model as the time scale when the perturbation theory with
respect to the electric field breaks down, and the \textrm{dc} response
changes from the linear-in-$E_{0}$ duration-independent\emph{\ }regime to a
nonlinear-in-$E_{0}$ and duration-dependent regime.\ In the experimental
situation described in Ref. \cite{Van+etal10}, a constant voltage between
two electrodes connected to the graphene was applied, $V=E_{0}L_{x}$, and
current-voltage characteristics ($I-V$) are measured within $\sim 1$ s,
which is a very large time scale compared with the ballistic flight time $%
T_{bal}=L_{x}/v_{F}$ for a finite flake length $L_{x}$. In typical
experiments, $L_{x}\sim 1\;\mathrm{\mu m}$, so that $T_{bal}\sim 10^{-12}\;%
\mathrm{s}$. To match our results with these conditions, our time $\Delta t$
should be replaced by some typical time scale that we call the effective
time duration $\Delta t_{eff}$. In the absence of the dissipation, the
transport is ballistic; in this case, considering a strip with a lateral
infinite width, we assume the ballistic flight time $T_{bal}$ to be the
effective time duration, $\Delta t_{eff}=T_{bal}$. In a realistic sample,
placed on a substrate, the effective time duration $\Delta t_{eff}$ can be
many times smaller than $T_{bal}$, because of charged impurities or the
structural disorder of the substrate. However, such an effective time $%
\Delta t_{eff}$ remains macroscopically large, so that Eq.~(\ref{4.51})
still holds. The external constant electric field can be considered as a
good approximation of the effective mean field as long as the field produced
by the induced current of created particles is negligible compared to the
applied field. Then $\langle j^{1}(t)\rangle ^{p}$ in Eq.~(\ref{4.41})
describes a regime where the current behaves as $j\sim V^{3/2}$. An
experimental observation of this $I-V$ was recently reported for
low-mobility samples (the case $\Delta t_{eff}$ $\ll T_{bal}$) \cite%
{Van+etal10}. {This implies the consistency restriction }$\Delta t\ll \Delta
t_{\mathrm{br}}=\Delta t_{\mathrm{st}}^{\mathrm{g}}/4\alpha $ \cite%
{GavGitY12}. Thus, there is a {window} in the parameter range of $E_{0}$ and
$\Delta t$ where the model with constant external field is consistent. For
example, let us assume that{\ }$\Delta t\sim ${$T_{bal}$}. It implies that $%
7\times 10^{-4}\,\mathrm{V}\ll V\ll 8\,\mathrm{V}$. These voltages are in
the range typically used in experiments with the graphene.

\section{Concluding remarks\label{S4}}

In the present article, we have revised vacuum instability effects in three
exactly solvable cases in QED with $t$-electric potential steps that have
real physical importance. These are the Sauter-like electric field, the
so-called $T$-constant electric field and exponentially growing and decaying
strong{\large \ }electric fields {\large \ }in a slowly varying regime.
Defining the slowly varying regime in general terms, we can observe the
existence of universal forms for the time evolution of vacuum effects caused
by strong electric field.{\large \ }Such universality appears when the
duration of the external field is sufficiently large in comparison to the
scale $\Delta t_{\mathrm{st}}=\left[ e\overline{E(t)}\right] ^{-1/2}$. In
this case the scale of{\large \ }the time varying for an external field and
leading contributions\textrm{\ to} vacuum mean values are macroscopic. Here,
we find\ universal approximate representations for the total density of
created pairs and{\large \ }vacuum means of current density and EMT
components that hold true for an arbitrary $t$-electric potential step
slowly varying with time. These representations do not require knowledge of
corresponding solutions of the Dirac equation; they have a form of simple
functionals of a given slowly varying electric field.{\large \ }We establish
relations of these representations with leading term approximations of
derivative expansion results. In fact, it is the possibility to adopt a
locally constant field approximation which makes an effect universal.{\large %
\ }These results allow one to formulate some semiclassical approximations
that are not restricted by smallness of differential mean numbers of created
pairs.{\large \ }We have\textbf{\ }tested the obtained representations in
the cases of exactly solvable $t$-electric potential steps.\textbf{\ }For
time instants $t$ close enough to the final time $t_{\mathrm{out}}$, $%
t\rightarrow t_{\mathrm{out}}$, the leading vacuum characteristics are
formed due to real pair production. One can say that{\large \ }we have
isolated global contributions that depend on the total history of an
electric field. Current density and EMT components of created pairs for a $T$%
-constant electric field can be easily extracted from the above-mentioned
representations. In such a way components of a\ Sauter-like\ field and
exponentially growing and decaying fields for $t\rightarrow t_{\mathrm{out}}$
are obtained for the first time. All these densities are growing functions
of the increment of the longitudinal kinetic momentum. However, their
explicit forms differ, in particular, since switching on and off conditions
of electric fields are different. It should also be noted that a universal
behavior of the vacuum mean current and EMT components was discovered for
time intervals, inside of which the electric field potential can be
approximated by a potential of a constant electric field. We see that for
such time{\large \ }intervals{\large \ }components of vacuum means of
current density and EMT can be divided into global and local contributions%
{\large . }Note that the global contributions depend on the effective time
duration of the electric field,{\large \ }$\Delta t$, and do not depend on
switching-off manner while the local contributions do not depend on the
interval $\Delta t$ and are functions of slowly varying electric field, $%
E\left( t\right) $. The global contributions define equations of state for
the matter field, which is a plasma of some kind of electron-positron
excitations created from vacuum. We show that local components of vacuum
mean EMT \ can be expressed via the one-loop effective Euler-Heisenberg
Lagrangian of the Dirac field and satisfy an equation of state for
electromagnetic field.

The reason for the universal behavior in the case under consideration is the
following: for total physical quantities as current density and EMT of
created pairs a large effective time of the field duration corresponds to a
large density of states that are occupied by created pairs if an electric
field is strong enough. One can guess that the universality under the
question is associated with the big state density that is a large parameter
in the slowly varying regime. Technically, we take into account only leading
terms with respect to these large parameter terms, whereas oscillation terms
are disregarded. In fact, using the approximation in question, we explicitly
show that the pair creation can be treated as a phase transition from the
initial vacuum to a plasma of electron-positron pairs.

\subparagraph{\protect\large Acknowledgements}

The work of S.P.G. and D.M.G. is supported by Tomsk State University
Competitiveness Improvement Program. The reported study of S.P.G. and D.M.G.
was partially supported by Russian Foundation for Basic Research (RFBR),
research project no. 15-02-00293a. D.M.G. is also supported by Grant No.
2016/03319-6, Funda\c{c}\~{a}o de Amparo \`{a} Pesquisa do Estado de S\~{a}o
Paulo (FAPESP), and permanently by Conselho Nacional de Desenvolvimento Cient%
\'{\i}fico e Tecnol\'{o}gico (CNPq), Brazil.

\appendix

\section{Vacuum instability in QED with $t$-electric potential steps\label%
{ApA}}

The nonperturbative approach to the $d=D+1$-dimensional model of Dirac
fields interacting with a strong $t$-electric potential steps is based on
the complete sets of exact solutions of the Dirac equation. Potentials $%
A^{\mu }\left( x\right) ,$ $x=(x^{\mu })=(x^{0}=t,\mathbf{r}),\;\mathbf{r}%
=(x^{i})$ of external electromagnetic fields\footnote{%
The greek indexes span the Minkowiski space-time, $\mu =0,1,\dots ,D$, and
the latin indexes span the Euclidean space, $l=1,\ldots ,D$. We use the
system of units where $\hslash =c=1$.
\par
{}} corresponding to $t$-electric potential steps are defined as%
\begin{equation}
A^{0}=0\,,\ \ \mathbf{A}\left( t\right) =(A^{1}=A_{x}\left( t\right) \,,\ \
A^{l}=0\,,\ \ l=2,...,D),\ A_{x}\left( t\right) \overset{t\rightarrow \pm
\infty }{\longrightarrow }A_{x}\left( \pm \infty \right) ,  \label{2.6}
\end{equation}%
where $A_{x}\left( \pm \infty \right) $ are some constant quantities, and
the time derivative of the potential $A_{x}\left( t\right) $ does not change
its sign for any $t\in \mathbb{R}.$ For definiteness, it is supposed that $%
\dot{A}_{x}\left( t\right) \leq 0\Longrightarrow A_{x}\left( -\infty \right)
>A_{x}\left( +\infty \right) .$ We stress that homogeneous electric fields
under consideration $\mathbf{E}\left( t\right) =\left( E_{x}\left( t\right)
,0,...,0\right) $ are switched off as $\left\vert t\right\vert \rightarrow
\infty $, $E_{x}\left( t\right) =-\dot{A}_{x}\left( t\right) =E\left(
t\right) \geq 0$,$\,E\left( t\right) \overset{\left\vert t\right\vert
\rightarrow \infty }{\longrightarrow }0.$

The Dirac equation reads%
\begin{eqnarray}
&&i\partial _{t}\psi \left( x\right) =H\left( t\right) \psi \left( x\right)
\,,\ \ H\left( t\right) =\gamma ^{0}\left( \boldsymbol{\gamma }\mathbf{P}%
+m\right) ,  \notag \\
&&\,P_{x}=-i\partial _{x}-U\left( t\right) ,\ \ \mathbf{P}_{\bot }=-i%
\boldsymbol{\nabla }_{\perp },\ \ U\left( t\right) =qA_{x}\left( t\right) \,,
\label{2.9}
\end{eqnarray}%
where $H\left( t\right) $ is the one-particle Dirac Hamiltonian, $\psi (x)$
is a $2^{[d/2]}$-component spinor, $[d/2]$ stands for the integer part of $%
d/2$, $m\neq 0$ is the electron mass, and the index $\perp $ stands for
components of the momentum operator that are perpendicular to the electric
field. Here, $\gamma ^{\mu }$ are the $\gamma $-matrices in $d$ dimensions
\cite{BraWey35}. The number of spin degrees of freedom is $J_{(d)}=2^{\left[
d/2\right] -1}$. We choose the electron as the main particle with the charge
$q=-e$, where $e>0$ is the absolute value of the electron charge.

The quantization of the Dirac field in the background under consideration is
based on the existence of solutions to the Dirac equation with special
asymptotics as $t\rightarrow \pm \infty $. For instance, we let the electric
field be switched on at $t_{\mathrm{in}}$ and switched off at $t_{\mathrm{out%
}}$, so that the interaction between the Dirac field and the electric field
vanishes at all time instants outside the interval $t\in \left( t_{\mathrm{in%
}},t_{\mathrm{out}}\right) $. We choose that before time $t_{\mathrm{in}}$
and after time $t_{\mathrm{out}}$ the spinors $\psi _{n}\left( x\right) $, $%
n=(\mathbf{p},\sigma )$, are states with a definite momentum $\mathbf{p}%
=\left( p_{x},\mathbf{p}_{\bot }\right) $ and spin polarization $\sigma
=(\sigma _{1},\sigma _{2},\dots ,\sigma _{\lbrack d/2]-1})$, $\ \sigma
_{s}=\pm 1$, and that they satisfy the following eigenvalue problems:%
\begin{eqnarray}
&&H\left( t\right) \ _{\zeta }\psi _{n}\left( x\right) =\ _{\zeta
}\varepsilon _{n}\ _{\zeta }\psi _{n}\left( x\right) \,,\ \ t\in \left(
-\infty ,t_{\mathrm{in}}\right] \,,\ _{\zeta }\varepsilon _{n}=\zeta
p_{0}\left( t_{\mathrm{in}}\right) \,,  \notag \\
&&H\left( t\right) \ ^{\zeta }\psi _{n}\left( x\right) =\ ^{\zeta
}\varepsilon _{n}\ ^{\zeta }\psi _{n}\left( x\right) \,,\ \ t\in \left[ t_{%
\mathrm{out}},+\infty \right) \,,\ ^{\zeta }\varepsilon _{n}=\zeta
p_{0}\left( t_{\mathrm{out}}\right) \,,  \notag \\
&&p_{0}\left( t\right) =\sqrt{\left[ p_{x}-U\left( t\right) \right] ^{2}+\pi
_{\perp }^{2}}\,,\;\pi _{\perp }=\sqrt{\mathbf{p}_{\perp }^{2}+m^{2}}.
\label{t4.ba}
\end{eqnarray}%
where the additional quantum number $\zeta =\pm $ labels states,
respectively. In these asymptotic states, $\zeta =+$ corresponds to free
electrons and $\zeta =-$ corresponds to free positrons.

In what follows, we consider two complete sets $_{\zeta }\psi _{n}\left(
x\right) $ and $\ ^{\zeta }\psi _{n}\left( x\right) $ of solutions of Dirac
equation (\ref{2.9}) (in- and out solutions respectively),%
\begin{eqnarray}
_{\zeta }\psi _{n}\left( x\right) &=&\left[ i\partial _{t}+H\left( t\right) %
\right] \gamma ^{0}\exp \left( i\mathbf{pr}\right) \ _{\zeta }\varphi
_{n}\left( t\right) v_{\chi ,\sigma },  \notag \\
\ ^{\zeta }\psi _{n}\left( x\right) &=&\left[ i\partial _{t}+H\left(
t\right) \right] \gamma ^{0}\exp \left( i\mathbf{pr}\right) \ ^{\zeta
}\varphi _{n}\left( t\right) v_{\chi ,\sigma }.  \label{t4.10}
\end{eqnarray}%
Here $v_{\chi ,\sigma }$ is a set of constant orthonormalized spinors, $%
\gamma ^{0}\gamma ^{1}v_{\chi ,\sigma }=\chi v_{\chi ,\sigma }$,$\ \chi =\pm
1$,\ $\ v_{\chi ,\sigma }^{\dag }v_{\chi ^{\prime },\sigma ^{\prime
}}=\delta _{\chi ,\chi ^{\prime }}\delta _{\sigma ,\sigma ^{\prime }}$. In
dimensions $d>3$, one can subject the spinors $v_{\chi }$ to some
supplementary conditions determining spin polarization $\sigma _{s}$ [in the
dimensions $d=2,3$ there are no spin degrees of freedom that are described
by the quantum numbers $\sigma $], and, together with the additional index $%
\chi $, provide a convenient parametrization of the solutions. The scalar
functions $\ _{\zeta }\varphi _{n}\left( t\right) $ and $^{\zeta }\varphi
_{n}\left( t\right) $ obey the second-order differential equation
\begin{equation}
\left\{ \frac{d^{2}}{dt^{2}}+\left[ p_{x}-U\left( t\right) \right] ^{2}+\pi
_{\perp }^{2}-i\chi \dot{U}\left( t\right) \right\} \left(
\begin{array}{c}
_{\zeta }\varphi _{n}\left( t\right) \\
^{\zeta }\varphi _{n}\left( t\right)%
\end{array}%
\right) =0\,.  \label{t3}
\end{equation}%
In the asymptotic regions%
\begin{eqnarray}
\ _{\zeta }\varphi _{n}\left( t\right) &=&\ _{\zeta }\mathcal{N}\exp \left[
-i\ _{\zeta }\varepsilon _{n}\left( t-t_{\mathrm{in}}\right) \right] \,,\ \
t\in \left( -\infty ,t_{\mathrm{in}}\right] \,,\   \notag \\
\ ^{\zeta }\varphi _{n}\left( t\right) &=&\ ^{\zeta }\mathcal{N}\exp \left[
-i\ ^{\zeta }\varepsilon _{n}\left( t-t_{\mathrm{out}}\right) \right] \,,\ \
t\in \left[ t_{\mathrm{out}},+\infty \right) \,,  \label{t4.1a}
\end{eqnarray}%
where$\ _{\zeta }\mathcal{N},\ ^{\zeta }\mathcal{N}$ are normalization
constants, and there exists an energy gap between the electron and positron
states. Since $\chi $ is not a physical quantum number [the spin operator $%
\gamma ^{0}\gamma ^{1}$ does not commute with the Dirac Hamiltonian (\ref%
{2.9}) in the case $m\neq 0$],\ we select the same $\chi $ for each $\
_{\zeta }\psi _{n}\left( x\right) $ and $\ ^{\zeta }\psi _{n}\left( x\right)
$. Solutions (\ref{t4.10}) are subject to the orthonormality conditions (the
standard volume regularization with a large spatial box of volume $V_{\left(
d-1\right) }$ is used). Then%
\begin{eqnarray}
&&\left( \ _{\zeta }\psi _{n},\ _{\zeta ^{\prime }}\psi _{n^{\prime
}}^{\prime }\right) =\delta _{n,n^{\prime }}\delta _{\zeta ,\zeta ^{\prime
}}\ ,\ \ \left( \ ^{\zeta }\psi _{n},\ ^{\zeta ^{\prime }}\psi _{n^{\prime
}}^{\prime }\right) =\delta _{n,n^{\prime }}\delta _{\zeta ,\zeta ^{\prime
}}\ ,  \notag \\
&&_{\zeta }\mathcal{N}\mathcal{=}\left[ 2p_{0}\left( t_{\mathrm{in}}\right)
q_{\mathrm{in}}^{\zeta }V_{\left( d-1\right) }\right] ^{-1/2}\,,\;\ ^{\zeta }%
\mathcal{N}=\left[ 2p_{0}\left( t_{\mathrm{out}}\right) q_{\mathrm{out}%
}^{\zeta }V_{\left( d-1\right) }\right] ^{-1/2}\   \label{t4.3}
\end{eqnarray}%
where $q_{\mathrm{in}/\mathrm{out}}^{\zeta }=p_{0}\left( t_{\mathrm{in}/%
\mathrm{out}}\right) -\chi \zeta \left[ p_{x}-U\left( t_{\mathrm{in}/\mathrm{%
out}}\right) \right] \,.$ The inner products $\left( \ _{\zeta ^{\prime
}}\psi _{n^{\prime }},\ ^{\zeta }\psi _{n}\right) $ are diagonal in quantum
numbers $n$ and $n^{\prime },$%
\begin{equation}
\left( \ _{\zeta ^{\prime }}\psi _{n^{\prime }},\ ^{\zeta }\psi _{n}\right)
=\delta _{n^{\prime },n}g\left( _{\zeta ^{\prime }}|^{\zeta }\right) ,\ \
g\left( ^{\zeta ^{\prime }}|_{\zeta }\right) =g\left( _{\zeta ^{\prime
}}|^{\zeta }\right) ^{\ast }.  \label{t4.5}
\end{equation}%
The corresponding diagonal matrix elements $g$ obey the unitarity relations%
\begin{equation}
\sum_{\varkappa }g\left( ^{\zeta }|_{\varkappa }\right) g\left( _{\varkappa
}|^{\zeta ^{\prime }}\right) =\sum_{\varkappa }g\left( _{\zeta }|^{\varkappa
}\right) g\left( ^{\varkappa }|_{\zeta ^{\prime }}\right) =\delta _{\zeta
,\zeta ^{\prime }}\,  \label{3.16.1}
\end{equation}%
and relate \textrm{in} and \textrm{out }solutions $\left\{ \ _{\zeta }\psi
_{n}\left( x\right) \right\} $ and $\left\{ \ ^{\zeta }\psi _{n}\left(
x\right) \right\} $ for each $n$,%
\begin{eqnarray}
^{\zeta }\psi _{n}\left( x\right) &=&g\left( _{+}|^{\zeta }\right) \
_{+}\psi _{n}\left( x\right) +g\left( _{-}|^{\zeta }\right) \ _{-}\psi
_{n}\left( x\right) \,,  \notag \\
_{\zeta }\psi _{n}\left( x\right) &=&g\left( ^{+}|_{\zeta }\right) \
^{+}\psi _{n}\left( x\right) +g\left( ^{-}|_{\zeta }\right) \ ^{-}\psi
_{n}\left( x\right) \,.  \label{t4.4}
\end{eqnarray}

Decomposing the Dirac operator $\hat{\Psi}(x)$ in the complete sets of in
and out solutions \cite{FGS,GavGT06},%
\begin{equation}
\hat{\Psi}\left( x\right) =\sum_{n}\left[ a_{n}(\mathrm{in})\ _{+}\psi
_{n}(x)+b_{n}^{\dag }(\mathrm{in})\ _{\_}\psi _{n}(x)\right] =\sum_{n}\left[
a_{n}(\mathrm{out})\ ^{+}\psi _{n}(x)+b_{n}^{\dag }(\mathrm{out})\ ^{-}\psi
_{n}(x)\right] \,,  \label{3.1}
\end{equation}%
we introduce \textrm{in} and \textrm{out }creation and annihilation Fermi
operators. Their nonzero anticommutation relations are,%
\begin{equation}
\lbrack a_{n}(\mathrm{in}),a_{m}^{\dag }(\mathrm{in})]_{+}=[a_{n}(\mathrm{out%
}),a_{m}^{\dag }(\mathrm{out})]_{+}=[b_{n}(\mathrm{in}),b_{m}^{\dag }(%
\mathrm{in})]_{+}=[b_{n}(\mathrm{out}),b_{m}^{\dag }(\mathrm{out}%
)]_{+}=\delta _{nm}\,.  \label{3.4}
\end{equation}%
In these terms, the Heisenberg Hamiltonian is diagonalized at $t\leq t_{%
\mathrm{in}}$ and $t\geq t_{\mathrm{out}}\ ,$%
\begin{eqnarray}
&&\widehat{\mathbb{H}}(t)=\sum_{n}\left\{ \ _{+}\varepsilon _{n}a_{n}^{+}(%
\mathrm{in})a_{n}(\mathrm{in})+\left\vert \ _{-}\varepsilon _{n}\right\vert
b_{n}^{+}(\mathrm{in})b_{n}(\mathrm{in})\right\} \,,\ \ t\leq t_{\mathrm{in}%
}\ ,  \notag \\
&&\widehat{\mathbb{H}}(t)=\sum_{n}\left\{ \;^{+}\varepsilon _{n}a_{n}^{+}(%
\mathrm{out})a_{n}(\mathrm{out})+\left\vert \ ^{-}\varepsilon
_{n}\right\vert b_{n}^{+}(\mathrm{out})b_{n}(\mathrm{out})\right\} \,,\ \
t\geq t_{\mathrm{out}}\ ,  \label{3.5}
\end{eqnarray}%
where the diverging c-number parts have been omitted, as usual. The initial $%
|0,$\textrm{$in$}$\rangle $ and final $|0,$\textrm{$out$}$\rangle $ vacuum
vectors, as well as many-particle \textrm{in} and \textrm{out }states, are
defined by%
\begin{eqnarray}
&&\ a_{n}(\mathrm{in})|0,\mathrm{in}\rangle =b_{n}(\mathrm{in})|0,\mathrm{in}%
\rangle =0,\ a_{n}(\mathrm{out})|0,\mathrm{out}\rangle =b_{n}(\mathrm{out}%
)|0,\mathrm{out}\rangle =0\,,  \notag \\
&&\ |\mathrm{in}\rangle =b_{n}^{+}(\mathrm{in})...a_{n}^{+}(\mathrm{in}%
)...|0,\mathrm{in}\rangle ,\ \ |\mathrm{out}\rangle =b_{n}^{+}(\mathrm{out}%
)...a_{n}^{+}(\mathrm{out})...|0,\mathrm{out}\rangle \,.  \label{3.6a}
\end{eqnarray}%
Using the charge operator one can see that $a_{n}^{\dag }$, $a_{n}$ are the
creation and annihilation operators of electrons, whereas $b_{n}^{\dag }$, $%
b_{n}$ are the creation and annihilation operators of positrons.

Transition amplitudes in the Heisenberg representation have the form $M_{%
\mathrm{in}\rightarrow \mathrm{out}}=\langle \mathrm{out}|\mathrm{in}\rangle
\,.$ In particular, the vacuum-to-vacuum transition amplitude reads $%
c_{v}=\langle 0,\mathrm{out}|0,\mathrm{in}\rangle .$ Relative probability
amplitudes of particle scattering, pair creation and annihilation are:%
\begin{eqnarray}
&&w\left( +|+\right) _{n^{\prime }n}=c_{v}^{-1}\langle 0,\mathrm{out}%
\left\vert a_{n^{\prime }}\left( \mathrm{out}\right) a_{n}^{\dagger }(%
\mathrm{in})\right\vert 0,\mathrm{in}\rangle =\delta _{n,n^{\prime
}}w_{n}\left( +|+\right) ,  \notag \\
&&w\left( -|-\right) _{n^{\prime }n}=c_{v}^{-1}\langle 0,\mathrm{out}%
\left\vert b_{n^{\prime }}\left( \mathrm{out}\right) b_{n}^{\dagger }(%
\mathrm{in})\right\vert 0,\mathrm{in}\rangle =\delta _{n,n^{\prime
}}w_{n}\left( -|-\right) \,,  \notag \\
&&w\left( +-|0\right) _{n^{\prime }n}=c_{v}^{-1}\langle 0,\mathrm{out}%
\left\vert a_{n^{\prime }}\left( \mathrm{out}\right) b_{n}\left( \mathrm{out}%
\right) \right\vert 0,\mathrm{in}\rangle =\delta _{n,n^{\prime }}w_{n}\left(
+-|0\right) \,,  \notag \\
&&w\left( 0|-+\right) _{nn^{\prime }}=c_{v}^{-1}\langle 0,\mathrm{out}%
\left\vert b_{n}^{\dagger }(\mathrm{in})a_{n^{\prime }}^{\dagger }(\mathrm{in%
})\right\vert 0,\mathrm{in}\rangle \,=\delta _{n,n^{\prime }}w_{n}\left(
0|-+\right) .  \label{3.15}
\end{eqnarray}

The \textrm{in} and \textrm{out }operators are related by linear canonical
transformations,%
\begin{equation}
a_{n}(\mathrm{out})=g(^{+}|_{+})a_{n}(\mathrm{in})+g(^{+}|_{-})b_{n}^{\dag }(%
\mathrm{in})\,,\ \ b_{n}^{\dag }(\mathrm{out})=g(^{-}|_{+})a_{n}(\mathrm{in}%
)+g(^{-}|_{-})b_{n}^{\dag }(\mathrm{in})\,.  \notag
\end{equation}%
These relations allow one to calculate the differential mean numbers of
electrons $N_{n}^{a}\left( \mathrm{out}\right) $ and positrons $%
N_{n}^{b}\left( \mathrm{out}\right) $ created from the vacuum state as%
\begin{eqnarray}
&&N_{n}^{a}\left( \mathrm{out}\right) =\left\langle 0,\mathrm{in}\left\vert
a_{n}^{\dagger }(\mathrm{out})a_{n}(\mathrm{out})\right\vert 0,\mathrm{in}%
\right\rangle =\left\vert g\left( _{-}\left\vert ^{+}\right. \right)
\right\vert ^{2},  \notag \\
&&N_{n}^{b}\left( \mathrm{out}\right) =\left\langle 0,\mathrm{in}\left\vert
b_{n}^{\dagger }(\mathrm{out})b_{n}(\mathrm{out})\right\vert 0,\mathrm{in}%
\right\rangle =\left\vert g\left( _{+}\left\vert ^{-}\right. \right)
\right\vert ^{2},\ N_{n}^{\mathrm{cr}}=N_{n}^{b}\left( \mathrm{out}\right)
=N_{n}^{a}\left( \mathrm{out}\right) .  \notag
\end{eqnarray}%
By $N_{n}^{\mathrm{cr}}$\ we denote the differential numbers of created
pairs . Relative probabilities (\ref{3.15}), the vacuum-to-vacuum transition
amplitude $c_{v}$, the probability for a vacuum to remain a vacuum $P_{v}$,
and the total number $N$ of pairs\ created from vacuum can be expressed via
the distribution $N_{n}^{\mathrm{cr}},$
\begin{eqnarray}
&&\left\vert w_{n}\left( +-|0\right) \right\vert ^{2}=N_{n}^{\mathrm{cr}%
}\left( 1-N_{n}^{\mathrm{cr}}\right) ^{-1},\;\left\vert w_{n}\left(
-|-\right) \right\vert ^{2}=\left( 1-N_{n}^{\mathrm{cr}}\right) ^{-1},
\notag \\
&&P_{v}=|c_{v}|^{2}=\prod\limits_{n}\left( 1-N_{n}^{\mathrm{cr}}\right) ,\ \
N^{\mathrm{cr}}=\sum_{n}N_{n}^{\mathrm{cr}}=\sum_{n}\left\vert g\left(
_{-}\left\vert ^{+}\right. \right) \right\vert ^{2}.  \label{vacprob}
\end{eqnarray}

The vacuum mean electric current, energy, and momentum are defined as
integrals over the spatial volume. Due to the translational invariance in
the uniform external field, all these mean values are proportional to the
space volume. Therefore, it is enough to calculate the vacuum mean values of
the current density vector $\langle j^{\mu }(t)\rangle $ and of the
energy-momentum tensor (EMT) $\langle T_{\mu \nu }(t)\rangle $, defined as
\begin{equation}
\langle j^{\mu }(t)\rangle =\langle 0,\mathrm{in}|j^{\mu }|0,\mathrm{in}%
\rangle ,\ \ \langle T_{\mu \nu }(t)\rangle =\langle 0,\mathrm{in}|T_{\mu
\nu }|0,\mathrm{in}\rangle \,.  \label{int1}
\end{equation}%
Here we stress the time dependence of mean values (\ref{int1}), which does
exist due to a time dependence of the external field. We recall for further
convenience the form of the operators of the current density and the EMT of
the quantum Dirac field,
\begin{align}
& j^{\mu }=\frac{q}{2}\left[ \overline{\hat{\Psi}}(x),\gamma ^{\mu }\hat{\Psi%
}\left( x\right) \right] \,,\quad T_{\mu \nu }=\frac{1}{2}(T_{\mu \nu
}^{can}+T_{\nu \mu }^{can})\,,  \notag \\
& T_{\mu \nu }^{can}=\frac{1}{4}\left\{ [\overline{\hat{\Psi}}(x),\gamma
_{\mu }P_{\nu }\hat{\Psi}\left( x\right) ]+[P_{\nu }^{\ast }\overline{\hat{%
\Psi}}(x),\gamma _{\mu }\hat{\Psi}\left( x\right) ]\right\} \,,  \notag \\
& P_{\mu }=i\partial _{\mu }-qA_{\mu }(x),\ \overline{\hat{\Psi}}(x)=\hat{%
\Psi}^{\dagger }\left( x\right) \gamma ^{0}.  \label{A1.0}
\end{align}

Note that the mean values (\ref{int1}) depend on the definition of the
initial vacuum, $|0,\mathrm{in}\rangle $ and on the evolution of the
electric field from the time $t_{\mathrm{in}}$ of switching it on up to the
current time instant $t$, but they do not depend on the further history of
the system. The renormalized vacuum mean values $\langle j^{\mu }(t)\rangle $
and $\langle T_{\mu \nu }(t)\rangle ,$ $t_{\mathrm{in}}$ $<t<$ $t_{\mathrm{%
out}}$ are sources in equations of motion for mean electromagnetic and
metric fields, respectively. In particular, complete description of the
backreaction is related to the calculation of these mean values for any $t$.

Mean values and probability amplitudes are calculated with the help of
different kinds of propagators. The probability amplitudes are calculated
using Feynman diagrams with the causal (Feynman) propagator%
\begin{equation}
S^{c}(x,x^{\prime })=i\langle 0,\mathrm{out}|\hat{T}\hat{\Psi}\left(
x\right) \hat{\Psi}^{\dagger }\left( x^{\prime }\right) \gamma ^{0}|0,%
\mathrm{in}\rangle c_{v}^{-1}\,,  \label{A1.3}
\end{equation}%
where $\hat{T}$ denotes the chronological ordering operation. A perturbation
theory (with respect to radiative processes) uses the so-called \textrm{in-in%
} propagator $S_{\mathrm{in}}^{c}(x,x^{\prime })$ and $S^{p}(x,x^{\prime })$
propagator,
\begin{equation}
S_{\mathrm{in}}^{c}(x,x^{\prime })=i\langle 0,\mathrm{in}|\hat{T}\hat{\Psi}%
\left( x\right) \hat{\Psi}^{\dagger }\left( x^{\prime }\right) \gamma ^{0}|0,%
\mathrm{in}\rangle ,\ \ S^{p}(x,x^{\prime })=S_{\mathrm{in}}^{c}(x,x^{\prime
})-S^{c}(x,x^{\prime }).  \label{A1.1}
\end{equation}

All the above propagators can be expressed via the \textrm{in} and \textrm{%
out }solution as follows:%
\begin{eqnarray}
S^{c}\left( x,x^{\prime }\right) &=&i\left\{
\begin{array}{c}
\sum\limits_{n}\ ^{+}\psi _{n}(x)\omega _{n}(+|+)\ _{+}\bar{\psi}%
_{n}(x^{\prime }),\ \ t>t^{\prime } \\
-\sum\limits_{n}\ _{-}\psi _{n}(x)\omega _{n}(-|-)\ ^{-}\bar{\psi}%
_{n}(x^{\prime }),\ \ t<t^{\prime }%
\end{array}%
\right. \,,  \label{3.22} \\
S_{\mathrm{in}}^{c}\left( x,x^{\prime }\right) &=&i\left\{
\begin{array}{c}
\sum\limits_{n}\ _{+}\psi _{n}(x)\ _{+}\bar{\psi}_{n}(x^{\prime }),\ \
t>t^{\prime } \\
-\sum\limits_{n}\ _{-}\psi _{n}(x)\ _{-}\bar{\psi}_{n}(x^{\prime }),\ \
t<t^{\prime }%
\end{array}%
\right. \,,\ \ S^{p}(x,x^{\prime })=\ -i\sum_{n}\,_{-}{\psi }%
_{n}(x)w_{n}\left( 0|-+\right) \,{_{+}\bar{\psi}}_{n}(x^{\prime })\,.
\label{3.25}
\end{eqnarray}

The mean values of the operator (\ref{A1.0}) are expressed via the latter
propagators as%
\begin{align}
& \langle j^{\mu }(t)\rangle =\mathrm{Re}\,\langle j^{\mu }(t)\rangle ^{c}+%
\mathrm{Re}\,\langle j^{\mu }(t)\rangle ^{p}\,,\ \ \langle j^{\mu
}(t)\rangle ^{c,p}=iq\left. \mathrm{tr}\left[ \gamma ^{\mu
}S^{c,p}(x,x^{\prime })\right] \right\vert _{x=x^{\prime }}\,,  \notag \\
& \langle T_{\mu \nu }(t)\rangle =\mathrm{Re}\,\langle T_{\mu \nu
}(t)\rangle ^{c}+\mathrm{Re}\,\langle T_{\mu \nu }(t)\rangle ^{p}\,,\ \
\langle T_{\mu \nu }(t)\rangle ^{c,p}=i\left. \mathrm{tr}\left[ A_{\mu \nu
}S^{c,p}(x,x^{\prime })\right] \right\vert _{x=x^{\prime }}\,,  \notag \\
& A_{\mu \nu }=1/4\left[ \gamma _{\mu }\left( P_{\nu }+P_{\nu }^{\prime \ast
}\right) +\gamma _{\nu }\left( P_{\mu }+P_{\mu }^{\prime \ast }\right) %
\right] \,.  \label{A1.4}
\end{align}%
Here $\mathrm{tr}$ stands for the trace in the $\gamma $-matrices indices
and the limit $x\rightarrow x^{\prime }$ is understood as follows:
\begin{equation*}
\mathrm{tr}[R(x,x^{\prime })]_{x=x^{\prime }}=\frac{1}{2}\left[
\lim_{t\rightarrow t^{\prime }-0}\mathrm{tr}[R(x,x^{\prime
})]+\lim_{t\rightarrow t^{\prime }+0}\mathrm{tr}\left[ R(x,x^{\prime })%
\right] \right] _{\mathbf{x=x}^{\prime }},
\end{equation*}%
where $R(x,x^{\prime })$ is any two-point matrix function.

The function $S^{p}(x,y)$ vanishes in the case of a stable vacuum. In this
case and only in this case $\langle j^{\mu }(t)\rangle =\mathrm{Re}\,\langle
j^{\mu }(t)\rangle ^{c}\ ,\ \ \ \langle T_{\mu \nu }(t)\rangle =\mathrm{Re}%
\,\langle T_{\mu \nu }(t)\rangle ^{c}.$

\end{document}